%% file: main.tex
\documentclass[11pt]{article}
\usepackage[margin=1in]{geometry}
\usepackage{booktabs,tabularx}
\usepackage{fullpage}
\usepackage{times}
\usepackage{url}
\usepackage{epsfig}
\usepackage{subfigure}
\usepackage{makecell}
\usepackage{graphicx}
\usepackage{titlesec}
\usepackage{multirow}
\usepackage{amsmath, amsthm, amssymb}

\titlespacing{\section}{0pt}{3mm}{1mm}
\titlespacing{\subsection}{0pt}{2mm}{0.5mm}
\titlespacing{\subsubsection}{0pt}{2mm}{0.8mm}

\usepackage{algorithm}
\usepackage{algorithmic}
\usepackage{listings}
\usepackage{times}
\usepackage{graphicx}
\usepackage{listings}
\usepackage{courier}
\usepackage{times}
\usepackage[T1]{fontenc}
\usepackage{cleveref}
\usepackage{amsmath}
\usepackage{caption}
\usepackage{amssymb}
\usepackage{latexsym}
\usepackage{bm}
\usepackage{color}
\usepackage{subscript}
\usepackage{multirow}
\usepackage{tgheros}
\usepackage{array}
\usepackage{verbatim}

\newcounter{Examplecount}
\makeatletter
 {\vspace{-3 pt} \begin{list}{}{%
  \setlength{\topsep}{0pt}\setlength{\partopsep}{0pt}\setlength{\parskip}{0pt}%
  \setlength{\parsep}{0pt}\setlength{\labelwidth}{0pt}%
  \setlength{\rightmargin}{0pt}\setlength{\leftmargin}{0pt}%
  \setlength{\labelsep}{0pt}%
  \obeylines\@vobeyspaces\normalfont\ttfamily%
  \item[]}}
 {\end{list}\vspace{-3 pt}\noindent}
\makeatother

\begin{document}
\clubpenalty=10000 
\widowpenalty = 10000

\date{\vspace{-5ex}}
\title{PlinyCompute: A Platform for High-Performance, Distributed, Data-Intensive Tool Development}
\maketitle

\begin{center} Jia Zou, R. Matthew Barnett, Tania Lorido-Botran, Shangyu Luo, Carlos Monroy \\ Sourav Sikdar, 
Kia Teymourian, Binhang Yuan, Chris Jermaine \\

\vspace{5 pt}
Rice University \\
Houston, Texas, USA
\vspace{5ex}
\end{center}
\input{code}


\input{abs}

\input{intro}

\input{pc_overview}

\input{object_model}

\input{tcap_optimization}

\input{experiments}

\input{rel_work}

\input{conc}

\bibliographystyle{abbrv}
\bibliography{pc}

\appendix
\input{example}

\input{allocation}

\input{pipelined_engine}

\input{implementation}

\end{document}

%% file: code.tex
\definecolor{dkgreen}{rgb}{0,0.6,0}
\definecolor{gray}{rgb}{0.5,0.5,0.5}
\definecolor{mauve}{rgb}{0.58,0,0.82}

\lstnewenvironment{code}
  {\lstset{
        aboveskip=10pt,
        belowskip=10pt,
        xleftmargin=15pt,        
        escapechar=!,
        mathescape=true,
        language=C++,
        basicstyle=\linespread{0.94}\ttfamily\footnotesize,
	    morekeywords={createSet, sendData, makeLambdaFromMember,
		execute, setInput, setOutput, makeLambda, makeLambdaFromMethod, 
		Handle, Vector, Map, pair, makeObjectAllocatorBlock,
	    push_back, Object, makeObject},
        keywordstyle=\color{blue}\ttfamily,
        commentstyle=\color{dkgreen}\ttfamily,			
        showstringspaces=true}
        \vspace{0pt}%
        \noindent\minipage{0.47\textwidth}}
  {\endminipage\vspace{0pt}}

\lstnewenvironment{codesmall}
  {\lstset{
        aboveskip=10pt,
        belowskip=10pt,
        xleftmargin=15pt,        
        escapechar=!,
        mathescape=true,
        language=C++,
        basicstyle=\linespread{0.94}\ttfamily\footnotesize,
	    morekeywords={createSet, sendData, makeLambdaFromMember,
		hash, join, APPLY, FILTER, execute, setInput, setOutput, makeLambda, makeLambdaFromMethod, 
		Handle, Vector, Map, pair, makeObjectAllocatorBlock,
        push_back, Object, makeObject},
        keywordstyle=\color{blue}\ttfamily,
        commentstyle=\color{dkgreen}\ttfamily,            			
        showstringspaces=true}
        \vspace{0pt}%
        \noindent\minipage{0.47\textwidth}}
  {\endminipage\vspace{0pt}}

%% file: abs.tex
\noindent
\textbf{Abstract}

\noindent
This paper describes \emph{PlinyCompute}, a system for development of
high-performance, data-intensive, distributed computing tools and libraries.
\emph{In the large}, PlinyCompute presents the programmer with a very high-level,
declarative interface, relying on automatic, relational-database style optimization to figure out how to stage
distributed computations.  However, \emph{in the small}, PlinyCompute
presents the capable systems programmer with a persistent object data model and API (the ``PC object model'') and associated memory management system
that has been designed from the ground-up for
high performance, distributed, data-intensive computing.
This contrasts with most other Big Data systems, which are constructed on top of the
Java Virtual Machine (JVM), and hence must
at least partially cede performance-critical concerns such as
memory management (including layout and de/allocation) and virtual
method/function dispatch to the JVM.
This hybrid approach---declarative in the large, trusting the programmer's ability
to utilize PC object model efficiently
in the small---results in a system that is ideal for the development of reusable, data-intensive tools and libraries.
Through extensive benchmarking, we show that implementing complex
objects manipulation and non-trivial, library-style computations 
 on top of PlinyCompute can result in a speedup of 2$\times$ to
 more than 50$\times$ or more compared to equivalent implementations on Spark.

%% file: intro.tex

\section{Introduction}
Big Data systems such as Spark \cite{zaharia2010spark} and Flink \cite{alexandrov2014stratosphere, carbone2015apache}
have effectively solved what we call the ``data munging'' problem.  That is, 
these systems do an excellent job supporting the rapid
and robust development of problem-specific,
distributed/parallel codes that transform a raw dataset into structured 
or semi-structured form, and then
extract actionable information from the transformed data.
But while existing Big Data systems offer
excellent support for data munging,
there is a class of application for which 
existing systems are 
used, but arguably are far less suitable:
as a platform 
for the development of high-performance codes, especially reusable
Big Data tools and libraries, by a capable
system programmer.

The desire to build new tools
on top of existing Big Data systems is understandable.  
The  developer of a distributed data processing tool must worry about data persistence, movement of
data to/from secondary storage, data
and task distribution, resource allocation, load balancing, fault tolerance, and many other factors.
While classical high-performance computing (HPC)
tools such as MPI \cite{gropp1996high} do not provide support for all of these concerns,
existing Big Data systems 
address them quite well.
As a result, many tools and libraries have been built on top of existing systems.  For example,
Spark supports
machine learning (ML) libraries Mahout \cite{mahout}, Dl4j \cite{dj4j}, 
and Spark \texttt{mllib} \cite{meng2016mllib}, linear algebra packages such as SystemML \cite{tian2012scalable, boehm2016systemml, ghoting2011systemml, boehm2014hybrid} and Samsara \cite{samsara}, and graph analytics with
GraphX \cite{gonzalez2014graphx} and GraphFrames
\cite{dave2016graphframes}.  Examples abound.

\vspace{5 pt}
\noindent
\textbf{PlinyCompute: A platform for high-performance, Big Data computing.}
However, we assert that if one were to develop a system purely for developing high-performance
Big Data codes
by a capable systems programmer,
it would not look like existing systems such as Spark, Flink,
DryadLinq~\cite{yu2008dryadlinq} and so on, 
which have largely 
been built using high-level programming languages and managed runtimes
such as the JVM and the .NET Common Language
Runtime (CLR).  Managed runtimes abstract away
most details regarding memory management
from the system designer, including memory allocation, deallocation,
reuse, and movement, as well as virtual function dispatch, 
object layout.
Since managing and utilizing memory is 
one of the most important factors determining big data system performance, reliance
on a managed environment can mean an order-of-magnitude increase in CPU cost for some computations~\cite{blackburn2006dacapo}.  
This cost may be acceptable if the person using the system
is a programmer uncomfortable with the basics of memory management who is
building an application to complete a specific data munging task.  
But it is unacceptable for high-performance tool or library
development by an expert. There have been notable efforts to engineer around the limitations of a managed environment and still provide
high performance---Spark's Dataset and
Dataframe abstractions come to mind---but such efforts are necessarily limited compared to
designing a Big Data system from the ground up around special-purpose
memory and object management system.

This paper is concerned with the design and implementation of
\emph{PlinyCompute}, a system for development of
high-performance, data-intensive, distributed computing codes, especially tools and libraries.
PlinyCompute, or PC for short, is designed to fill the gap between
HPC softwares such as OpenMP \cite{dagum1998openmp} and MPI \cite{gropp1996high}, which provide little direct support for
managing very large datasets, and dataflow platforms such as Spark and Flink, which 
may give up significant performance through their reliance on a managed runtime to handle
memory management (including layout and de/allocation) and key computational considerations
such as virtual function dispatch. 

\vspace{5 pt}
\noindent
\textbf{Core design principle: Declarative in the large, high-performance in the small.}
PC is unique in that \emph{in the large}, 
it presents the programmer with a very high-level,
declarative interface, relying on automatic, 
relational-database style optimization \cite{chaudhuri1998overview} to figure out how to stage
distributed computations.  
PC's declarative interface is higher-level than other Big Data systems
such as Spark and Flink, in that decisions such as choice of join ordering and which
join algorithms to run are
totally under control of the system. 
This is particularly important for tool and library development because the same tool should run well regardless of the data
it is applied to---the classical idea of \emph{data independence} in database system design \cite{stonebraker1990third}.
A relatively naive library user cannot be expected to tune a library implementation of an algorithm to run
well on his or her particular dataset, and yet with existing systems, this sort of tuning
is absolutely necessary.  For example, we find
that a high quality LDA implementation\footnote{LDA \cite{blei2003latent} is a popular text mining algorithm.}
on top of Spark is around $25\times$ slower than the algorithmically equivalent LDA
implementation on top of PC.  Through careful, dataset-specific tuning (including choosing specific join algorithms and
forcing pipelining of certain results) it is possible to get that gap down to $2.5\times$.  But this requires modification of the
code itself, which is beyond the vast majority of end-users.

In contrast, \emph{in the small}, PlinyCompute presents a capable programmer with a
persistent object data model and API (the ``PC object model'') and associated memory management system
designed from the ground-up for
high performance.
All data processed by PC are managed by
the PC object
model, which is exclusively responsible for PC data layout and within-page memory management.  
The PC object model is tightly coupled with
PC's execution engine, and has been specifically designed for
efficient distributed computing.  
All dynamic PC \texttt{Object} allocation is \emph{in-place}, directly on a page, obviating
the need for PC \texttt{Object} serialization and deserialization before data are transferred to/from storage or over a network.
Further, PC gives a programmer fine-grained control of the systems
memory management and PC \texttt{Object} re-use policies.

This hybrid approach---declarative and yet trusting the programmer
to utilize PC's object model effectively
in the small---results in a system that is ideal for the 
development of data-oriented tools and libraries.

The system consists of following components: 

\begin{itemize}
\item The \textbf{PC object model}, which is a toolkit for building high-performance, persistent data structures that can be 
processed and manipulated by PC.  

\item The \textbf{PC API and TCAP compiler}.  In the large, PC codes
  are declarative and look a lot like classical relational calculus
  \cite{codd1971data}.  For example, to specify a join over five sets
  of objects, a PC programmer does not build a join directed acyclic
  graph (DAG) over the five inputs, as in a standard
dataflow platform.  Rather, a programmer 
supplies two \emph{lambda term construction functions}: one that constructs a lambda term describing the selection
predicate over those five input sets, 
and a second that constructs a lambda term describing the relational projection over those five sets
using the same API.  These lambda terms are constructed using PC's built-in lambda abstraction families as well as higher-order composition functions.
 PC's TCAP compiler 
accepts such a specification, and compiles it into a functional, domain specific language called \emph{TCAP} that implements
the join.  Logically, TCAP operates over
sets of columns of PC \texttt{Object}s. 

\item The \textbf{execution engine}, which is a distributed query processing
  system for big data analytics. It consists of an optimizer for TCAP
  programs and a high-performance, distributed, vectorized TCAP
  processing engine.  
The TCAP processing engine has been designed to work closely with the PC object model to
minimize memory-related costs during computation.

\item \textbf{Various distributed services}, which include a catalog
  manager serving system meta-data, and a distributed storage manager.

\end{itemize}

\vspace{5 pt}
\noindent
\textbf{Our contributions.}
Taken together, these components allow a competent system programmer to write exceedingly high-performance distributed codes.
In this paper, we describe the design and implementation of PlinyCompute.  Currently, PC exists as a prototype system, consisting of around
150,000 lines of C++ code, with a much smaller amount of Prolog code. We experimentally show the performance benefits of the PC object model, 
demonstrating how even simple data transformations are much faster using the PC object model compared to similar computations within the 
Apache ecosystem.
In keeping with PC being targeted at high-performance
library and tool development,
we benchmark several library-style softwares written on top of PC.  We begin with a small domain specific language
for distributed linear algebra that we implemented on top of PC, called \texttt{LilLinAlg}.  \texttt{LilLinAlg} was implemented in about six weeks by a developer
who had no knowledge of PC at the outset, with the goal of demonstrating PC's suitability as a tool-development platform.  
We show that \texttt{LilLinAlg} has better performance than other systems that have long been under development
within the Apache ecosystem.   
We also benchmark the efficiency of manipulating complex objects, and
several standard machine learning codes written on top of PC.  

\vspace{5 pt}
\noindent
\textbf{Roadmap.} We first present an overview of PC runtime and the key components
: the PC object model, the lambda calculus that
forms the basis of PC API, 
and TCAP and PC's execution engine. Then we discuss PC object model
and TCAP optimization
in more detail in Section~\ref{sec:ObjectModel}  and
Section~\ref{sec:optimizer} respectively. In
Section~\ref{sec:exp}, we present a thorough 
evaluation and demonstrates that PlinyCompute outperforms alternatives
in building non-trivial, library-style computations  and manipulating
complex objects. Finally, we discuss related works in
Section~\ref{sec:survey} and summarize the paper in Section~\ref{sec:conc}.

%% file: pc_overview.tex
\section{Overview of PlinyCompute}

The PC software consists of 
(1) the PC object model, (2) the PC API and TCAP compiler (TCAP is a domain-specific language executed by PC),
(3) the execution engine, and (4) various distributed services. In the next few sections of the paper, we discuss the first three software components in detail.

When PC runs on a distributed cluster it has a \emph{master node} and one or more \emph{worker nodes}.
Running on the master node are the managers for the various distributed services provided by PC, primarily 
the \emph{catalog manager} and the \emph{distributed storage manager}.  Also running on the master
node is the software responsible for powering the distributed execution engine: the \emph{TCAP optimizer} and
the \emph{distributed query scheduler}.  
When a user of PC requests to execute a graph of computations, the
computations are compiled into a TCAP program on the user's process, then optimized
by the master node's TCAP optimizer and executed by the distributed query scheduler.

Each worker node runs two processes: the \emph{worker front-end process} and the \emph{worker backend process}.
Dual processes are used because the backend process
executes potentially unsafe native user code.
If
user code happens to crash the worker backend process, the worker 
front-end process can re-fork the worker
backend process.  

The worker front-end process interfaces with the master node, providing a local catalog manager and a local storage server (including
a local buffer pool)
and crucially, it acts as a proxy, forwarding requests to perform various computations to the worker backend process, where
computations are actually run.

\section{Overview of the Object Model}

There is growing evidence that the CPU costs associated with manipulating data, especially data (de-)serialization and memory 
(de-)allocation,  
dominate the time needed to complete typical big data processing tasks
\cite{ousterhout2015making, shi2015clash, SikdarSoCC2017}.
To avoid these potential pitfalls while at the same time giving the user a high degree of flexibility,
PC requires programmers to store and manipulate data using the \emph{PC object model}.
The PC object model is an API for storing and manipulating objects, and has been co-designed with PC's memory management system and execution engine to provide
maximum performance.  

In PC's C++ binding, individual PC \texttt{Object}s correspond to C++ objects, and so the C++ compiler specifies the memory layout.
However, where PC \texttt{Object}s are stored in RAM and on disk, and how they are allocated and deallocated, when and where they are moved, is
tightly controlled by PC itself.

The PC object model provides a fully object-oriented interface, and yet manages to avoid many of the costs associated with complex object manipulation
by following the \emph{page-as-a-heap} principle.  
All PC \texttt{Object}s are allocated and manipulated in-place, on a system-
(or user-) allocated page.  There is
no distinction between the in-memory representation of data and the on-disk (or in-network) representation of
data. Thus there is no (de-)serialization cost to move data to/from disk and network, and memory management costs are very low. Depending upon choices made by the
programmer, ``deallocating'' a page of objects
may mean simply unpinning the page and allowing it to be returned to the buffer pool, where it will be recycled and written over with a new set of objects.  
In computer systems design, this is often referred to as
\emph{region-based allocation} \cite{tofte1997region,
  grossman2002region}, and is often the fastest way to manage
memory. 

To illustrate the use of the PC object model from a user's perspective,
imagine that we wish to perform a computation over a number of feature vectors.  
Using the PC object model's C++ binding, we might represent each data point using the 
\texttt{DataPoint} class:

\begin{codesmall}
class DataPoint : public Object {
public:
	Handle <Vector <double>> data;
};
\end{codesmall}

\noindent
To create and load such data into a PC compute cluster, we might write the following code:

\begin{codesmall}
makeObjectAllocatorBlock (1024 * 1024);
Handle <Vector <Handle <DataPoint>>> myVec = 
     makeObject <Vector <Handle <DataPoint>>> ();
Handle <DataPoint> storeMe = makeObject <DataPoint> ();
storeMe->data = makeObject <Vector <double>> ();

for (int i = 0; i < 100; i++) 
     storeMe->data->push_back (1.0 * i);

myVec->push_back (storeMe);
pcClient.createSet <DataPoint> ("Mydb", "Myset");
pcClient.sendData <DataPoint> ("Mydb", "Myset", myVec);
\end{codesmall}

\noindent
Here, the programmer starts out by creating a one megabyte \emph{allocation block} where all new objects will be written,
and then allocates data directly to that allocation block via a call to \texttt{makeObject ()}.  Each call to  \texttt{makeObject ()}
returns a PC's pointer-like object, called a \texttt{Handle}.  PC \texttt{Handle}s use offsets rather than an absolute memory
addresses, so they can be moved from process to process and remain valid.  

When the data are dispatched via \texttt{sendData ()},
the occupied
portion of the allocation block is transferred in its entirety with
no pre-processing and zero CPU cost, aside from the cycles required to perform the data transfer.  
This illustrates the principle of \emph{zero cost data movement}.

Object allocation and deallocation is handled by PC object model.
If the next line of code executed were:
\texttt{
myVec = nullptr;}
then all of the memory associated with the \texttt{Vector} of \texttt{DataPoint} objects would be automatically
freed, since PC \texttt{Object}s are reference counted.  This can
be expensive, however, since the PC \texttt{Object} infrastructure must traverse a potentially large graph of \texttt{Handle} objects to perform the deallocation.  
Recognizing that low-level data manipulations dominate big data
compute times \cite{ousterhout2015making, shi2015clash}, PlinyCompute gives a programmer nearly complete control
over most aspects of memory management.

If the 
programmer had instead used: 

\begin{code}
storeMe->data = makeObject <Vector <double>> (ObjectPolicy :: noRefCount);
\end{code}

\noindent then the memory associated with \texttt{storeMe->data} would
not be reference counted, and hence not reclaimed when no longer
reachable.  

This may mean lower memory utilization,
but the benefit is nearly zero-CPU-cost memory management within the block.
PC gives the developer the ability to manage the tradeoff.
This illustrates another key principle behind the design of
PlinyCompute: \emph{Since PC is targeted towards tool and library
  development, PC assumes the programmer is capable.  In
  the small, PC gives the programmer all of the tools s/he needs to make things fast}.

Finally, we note that the PC object model is not used exclusively or
even primarily for application programming.  The PC object model
is used \emph{internally}, integrated with PC's execution engine as well.
For example, aggregation is implemented using PC's built-in
\texttt{Map} class.  Each thread maintains
a \texttt{Map} object that the thread aggregates its local data to; those are
merged into maps that are sent to various workers around the PC
cluster.  All sends and receives of these \texttt{Map} objects happen
without (de-)serialization.

\section{PlinyCompute's Lambda Calculus}
\label{sec:lambda}
A PC programmer specifies a distributed query graph by providing a graph of high-level computations over sets of data---those data
may either be of simple types, or they may be
PC \texttt{Object}s. 

The PC toolkit consists of a set of
computations 
that are not unlike the operations provided by systems such as Spark and Flink, though they are less numerous and generally higher-level:
\texttt{SelectionComp} (equivalent to relational selection and projection), \texttt{JoinComp} (equivalent to a join of arbitrary arity and arbitrary predicate), 
\texttt{AggregateComp} (aggregation), \texttt{MultiSelectComp} (relational selection with a set-valued projection function) and a few others.  
Each of these is an abstract type descending from PC's \texttt{Computation} class.

Where PC differs from other systems is that a programmer customizes these computations by writing code that composes together various C++ codes 
using a 
domain-specific lambda calculus.
For example, to implement a \texttt{SelectionComp} over PC \texttt{Object}s of type \texttt{DataPoint}, a programmer
must implement the lambda term construction function
\texttt{getSelection (Handle <DataPoint>)} and \texttt{getProject ion(Handle <DataPoint>)} which returns a lambda term
describing how \texttt{DataPoint} objects
should be processed.

Novice PC programmers sometimes incorrectly assume that the lambda construction functions operate on the data themselves, and
hence are called once for every data object in an input set---for example, 
that
\texttt{getSelection ()} would be repeatedly invoked to filter each \texttt{DataPoint} in an input set.  
This is incorrect, however.
A programmer is not supplying a computation over input data; rather, a programmer is supplying an expression in the lambda calculus that 
specifies \emph{how to construct the computation}.

To construct statements in the lambda calculus, PC supplies a programmer with a set of built-in \emph{lambda abstraction} 
families \cite{miller1991logic}, as 
well as a set of \emph{higher-order functions} \cite{chen1993hilog}
that take as input one or more lambda terms, and returns a new lambda term.  Those built-in lambda abstraction families 
include:

\begin{enumerate}

\vspace{-5pt}
\item \texttt{makeLambdaFromMember ()}, which returns 
a lambda abstraction taking as input a \texttt{Handle} to a PC \texttt{Object}, and returns a function returning one of the pointed-to object's member variables;

\vspace{-5pt}
\item 
\texttt{makeLambdaFromMethod ()}, which is similar, but returns a function calling a method on the pointed-to variable;

\vspace{-5pt}
\item \texttt{makeLambda ()}, which returns a function calling
a native C++ lambda;

\vspace{-5pt}
\item \texttt{makeLambdaFromSelf ()}, which returns the identity function.

\end{enumerate}

\vspace{-5pt}
\noindent
When writing a lambda term construction function, a PC programmer uses these families to create lambda abstractions that
are customized to a particular task.
The higher-order functions provided are used to compose lambda terms, and
include functions corresponding to:

\begin{enumerate}
\vspace{-5pt}
\item
The standard boolean comparison operations: \texttt{==}, \texttt{>}, \texttt{!=}, etc.;

\vspace{-5pt}
\item
The standard boolean
operations: \texttt{\&\&}, \texttt{||}, \texttt{!}, etc.;

\vspace{-5pt}
\item
The standard arithmetic operations: \texttt{+}, \texttt{-}, \texttt{*}, etc.  
\end{enumerate}

For an example of all of this, consider performing a join over three
sets of PC \texttt{Object}s stored in the PC cluster.  
Joins are specified in PC 
by implementing a \texttt{JoinComp} object. One of the methods that
must be overridden to build a specific join is \texttt{JoinComp :: getSelection ()}
which returns a lambda term
that specifies how to compute if a particular combination of input objects is accepted by the join.  Consider the following
\texttt{getSelection ()} for a three-way join over objects of type \texttt{Dept}, \texttt{Emp}, and \texttt{Sup}:

\begin{codesmall} 
Lambda <bool> getSelection (Handle <Dep> arg1, Handle <Emp> arg2, Handle <Sup> arg3) {
	return makeLambdaFromMember (arg1, deptName) == 
	       makeLambdaFromMethod (arg2, getDeptName) &&
	       makeLambdaFromMember (arg1, deptName) == 
               makeLambdaFromMethod (arg3, getDept);   }
\end{codesmall}

\noindent
This method creates a lambda term taking three arguments \texttt{arg1, arg2, arg3}.  This lambda terms describes a computation that
checks to see if \texttt{arg1->deptName} is the same as the value returned from \texttt{arg2->getDeptName ()}, and
that \texttt{arg1->deptName} is the same as the value returned from \texttt{arg3}\-\texttt{->getDept ()}. 
Note that the programmer does \emph{not} specify an ordering for the joins, and does \emph{not} specify specific join algorithms or variations.  Rather, PC
analyzes the lambda term returned by \texttt{getSelection ()} and
makes such decisions automatically.

In general, a programmer can choose to expose the details of a computation to PC, by making extensive use of PC's lambda
calculus, or not.  A programmer could, for example, hide the entire selection predicate within a native C++ lambda.
If the programmer chose to do this, PC would be unable to optimize the compute plan---the system relies on the willingness of the 
programmer to expose intent via the lambda term construction function.

\vspace{5pt}
A complete example of using PC APIs, which is based on the lambda
calculus as described in this section, can be found in
Section~\ref{sec:example} in the Appendix.

\begin{figure}[t]
  \begin{center}
    \includegraphics[width=4in]{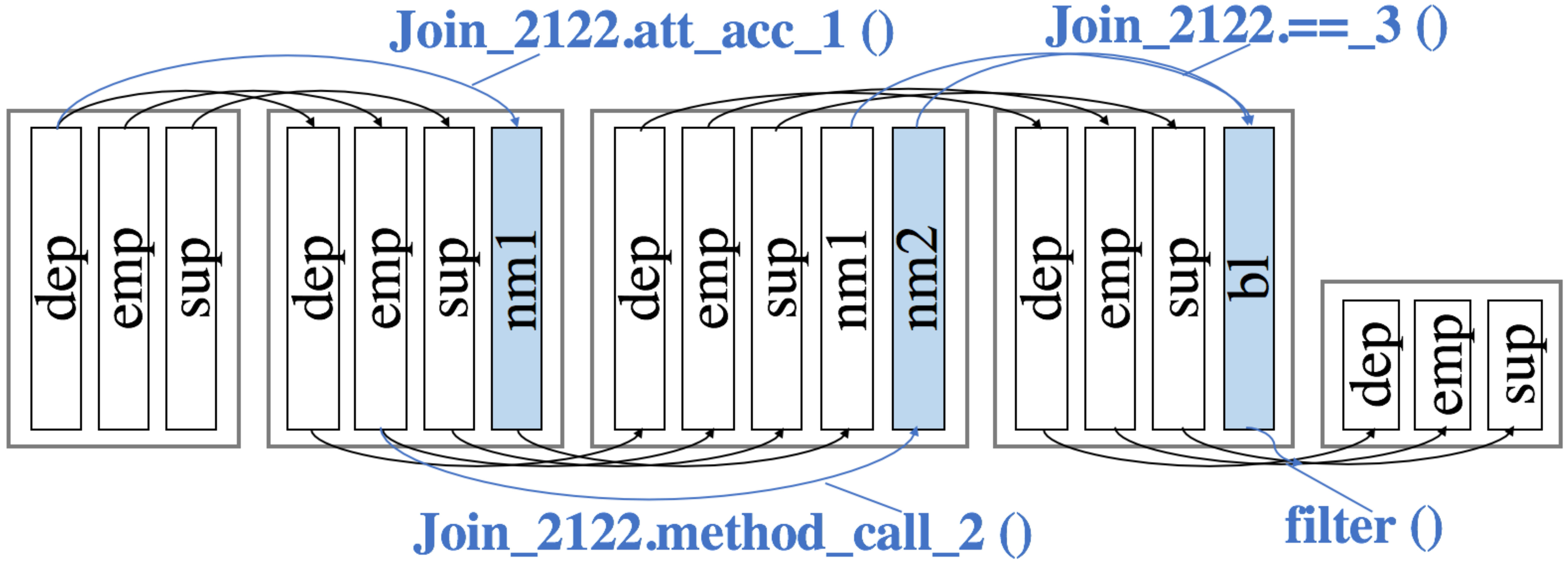}
  \end{center}
  \caption{Execution of the first four stages of a pipeline constructed from the example TCAP program.  The first two stages extract new vectors from 
existing vectors of PC \texttt{Object}s, first via a call to \texttt{Join\_2122.att\_acc\_1 ()}, which extracts \texttt{Dep.deptName} from each item in the vector
\texttt{dep} 
of \texttt{Dep} objects, producing a new vector called \texttt{nm1}. Then, second
via a call to \texttt{Join\_2122.method\_call\_2 ()}, which invokes \texttt{Dep :: getDeptName ()} on each of the \texttt{Emp} objects
in the vector \texttt{emp}, producing a new vector called \texttt{nm2}.  A bit vector \texttt{bl} is formed by checking the equality of those two 
vectors via a call to \texttt{Join\_2122.==\_3 ()}, and finally all of the vectors are filtered.}
  \label{fig:TCAP}
\end{figure}

\section{PlinyCompute's Execution Engine}
\label{sec:engine}
PC's execution engine is tasked with optimizing and executing TCAP programs (pronounced ``tee-cap'').  
PC's TCAP compiler
calls the various user-supplied lambda term construction functions for each of the \texttt{Computation} objects in a
user-supplied graph of computations,
and compiles all of those lambda terms into a DAG of small, atomic operations---a TCAP program.
A TCAP program is fully optimizable, using many standard techniques from relational query execution, as we
will discuss in Section:~\ref{sec:optimizer}.
In this section, we discuss how TCAP programs are executed by PC.

\subsection{Vectorized or Compiled?}
\label{sec:vectorized}

Volcano-style, record-by-record iteration \cite{graefe1990encapsulation} has fallen out of favor over the last decade, largely replaced by
two competing paradigms for processing data
in high-performance, data-oriented computing.  The first is \emph{vectorized} processing \cite{abadi2009column, boncz2005monetdb, zukowski2005monetdb, idreos2012monetdb}, where a column of values are pushed
through a simple computation in sequence, so as to play to the strength of a modern CPU, with few cache misses and no virtual
function calls.  The second is \emph{code generation} \cite{neumann2011efficiently, nagel2014code, bress2017generating, klonatos2014building, ahmad2009dbtoaster}, where a system analyzes the computation
and then generates code---either C/C++ code, or byte code for a framework such as LLVM \cite{lattner2004llvm, lattner2002llvm}.

While PlinyCompute certainly leverages ideas from both camps, we argue that the ``vectorized vs. generated'' argument is relevant mostly for 
relational systems with a data-oriented, domain-specific language (such as SQL).  
The data manipulations directly specified by an SQL programmer are likely to be limited, 
consisting of comparison between attributes, simple arithmetic, and logical operations.
Applying classical vectorization to PC,
which requires an execution plan to be constructed consisting entirely of calls to a toolkit of
vector-based operations shipped with the system, is
unrealistic when most/all computations are over user-defined
types. Further, generating LLVM code for complex operations over user-defined types 
in a high-level language 
is akin to writing a full-fledged compiler.

PC uses a hybrid approach, where the PC execution engine is vectorized, but the code for the individual vectorized operations (called \emph{pipeline stages})
is fully compiled.  PC's C++ binding relies on template metaprogramming (see Section 5.3)
to convert the user-supplied lambda terms 
(see Secion 4) into efficient pipeline stages over vectors of PC
\texttt{Object}s or simple types.

The operations in this DAG are then
optimized (that is, operations are automatically
re-ordered to form an optimal plan) using classical relational
methods \cite{chaudhuri1998overview, graefe1995cascades, jarke1984query}. After optimization, the pipeline stages are fit together to produce a set of interconnected pipelines.  Input data are broken into lists of 
data vectors (called, appropriately, \emph{vector lists}), and fed into the various pipelines.
Optimization of the DAG of pipeline stages
is possible because the programmer expresses intent via the lambda calculus \cite{barendregt1984lambda, moggi1989computational}.
Thus, PC's hybrid approach is vectorized, 
but it is \emph{also} compiled---the opaque C++ user code is compiled into pipeline stages that are 
assmebled into an optimized plan.

\subsection{TCAP and Vectorized Execution} \label{sec:vectorized}

As mentioned in Section \ref{sec:vectorized}, PC's vectorized execution engine repeatedly pushes 
so-called \emph{vector lists} through a pipeline that runs a series of
pipeline stages.  Each pipeline stage
takes as input a vector list,
and produces a 
new vector list that consists of zero or more vectors from the input
vector list, as well as one or more new vectors, appended at the end
of the list.

Pipeline stages are constructed in such a way that the overhead of a
vritual function call can be amortized on a vector list of objects, aside from any virtual function calls that may be
present (explicitly or perhaps implicitly in the form of memory
management) in the user's code.
The number of objects in a vector can be tuned to fit the L1 or
L2 cache size, depending on the object sizes and virtual function call
overhead. All vectors in a vector list should have the same number of objects.

The TCAP language describes both the pipeline stages required to perform a PC computation, as well as the schema for each of the vector lists that
will be produced during the PC computation, and how each of the pipeline stages adds or removes vectors from the vector lists that are pushed through
the computation.

To see how this works through an example, consider a variant of the \texttt{getSelection ()}:

\begin{codesmall} 
Lambda <bool> getSelection (Handle <Dep> arg1, Handle <Emp> arg2, Handle <Sup> arg3) {
	return makeLambdaFromMember (arg1, deptName) == 
	       makeLambdaFromMethod (arg2, getDeptName);  }
\end{codesmall}

\noindent PC compiles the lambda term resulting from a call to \texttt{getSelection ()} into the following TCAP code:

\begin{codesmall}
WDNm_1(dep,emp,sup,nm1) <= APPLY(In(dep), In(dep,emp,sup), 'Join_2212', 'att_acc_1', 
  [('type', 'attAccess'), ('attName', 'deptName')]);

WDNm_2(dep,emp,sup,nm1,nm2) <= APPLY(WDNm_1(emp), WDNm_1(dep,emp,sup,nm1), 'Join_2212',
  'method_call_2', [('type', 'methodCall'), ('methodName', 'getDeptName')]);

WBl_1(dep,emp,sup,bl) <= APPLY(WDNm_2(nm1,nm2),WDNm_2(dep,emp,sup), 'Join_2212', '==_3', 
  [('type', 'equalityCheck')]);

Flt_1(dep,emp,sup) <= FILTER(WBl_1(bl), WBl_1(dep,emp,sup), 'Join_2212', []);
\end{codesmall}

\noindent
These four TCAP statements correspond to a pipeline of four stages,
as shown above in Figure \ref{fig:TCAP}.

This particular TCAP code begins with an \texttt{APPLY} operation, which is a five-tuple, consisting of: (1) the vector list and constituent
vector(s) for the \texttt{APPLY} to operate on, (2) the vector(s) from that vector list to copy
from the input to the output, (3) the name of the computation that the operation was compiled from, (4) the name of the compiled code (pipeline stage)
that the operation is to execute, plus (5) a key-value map that stores specific information about the operation that may be used 
later during optimization.

Specifically, in this case, the first \texttt{APPLY} in the TCAP computation describes the following.  It describes a pipeline stage that
takes as input 
a vector list called \texttt{In}, which is made of the constituent vectors, referred to using 
the names \texttt{dep}, \texttt{emp}, and \texttt{sup}.
To produce the output vector list (called \texttt{WDNm\_1}), the vectors
\texttt{dep}, \texttt{emp}, and \texttt{sup} should be simply copied (via a shallow copy) from \texttt{In}.
In addition, the compiled code referred to by \texttt{Join\_2212.att\_acc\_1} will be executed via a vectorized application to the input
vector \texttt{dep}.  The result will then be put into a new vector called 
\texttt{WDNm\_1.nm1}.
The resulting vector list (consisting of the vectors shallow copied from the input as well as the new vector \texttt{WDNm\_1.nm1})
will be called \texttt{WDNm\_1}.  

The TCAP program also
specifies that \texttt{WDNm\_1} is processed by \texttt{APPLY}ing the
method call \texttt{getDeptNa me()} on the attribute \texttt{emp}; this is 
done via application of the compiled code referred to by
\texttt{Join\_2212.met hod\_call\_2}.  
The vectors \texttt{dep}, \texttt{emp}, \texttt{sup}
and \texttt{nm1} are simply shallow-copied to the output vector list.

After this an equality check is performed to create \texttt{WBl\_1.bl}
(a vector of booleans) and then the result is filtered based upon this
column.

Note that in each TCAP operation, the key-value map is only informational and does not affect its execution.  However, this
information can be vital during optimization.  For example, as we will discuss in Section~\ref{sec:optimizer}, 
multiple calls to the same method can be detected using information stored in TCAP key-value maps, and the redundant calls eliminated.

\subsection{Template Metaprogramming}
\label{sec:template}
In PC,
each vectorized pipeline stage (such as \texttt{Join\_2212.att\_acc\_1}) is executed as fully-compiled native code, with no virtual function
calls.
In PC's C++ binding, this is accomplished by using the C++ language's extensive \emph{template metaprogramming} 
capabilities \cite{josuttis2012c++}.  Templates are the C++ language's way of providing generics functionality.
When a C++ template class or
function is instantiated with a type, 
the C++ compiler actually generates optimized native code for that specific new type, at compile time.  
This is quite different from languages
such as Java, that must typically rely on slow virtual
function calls in order to implement generics.

To see how template metaprogramming is used by PC, consider
the TCAP operation from our example:

\begin{codesmall}
WBl_1(dep,emp,sup,bl) <= 
   APPLY(WDNm_2(nm1,nm2),WDNm_2(dep,emp,sup), 'Join_2212', '==_3', '');
\end{codesmall}

\noindent
Here, the pipeline stage \texttt{Join\_2212.==\_3} that is specified by the \texttt{APPLY} operation
actually refers to a function generated as a by-product of the
programmer using PC's \texttt{==} operation
in the line of code

\begin{codesmall} 
	return makeLambdaFromMember (arg1, deptName) == 
	       makeLambdaFromMethod (arg2, getDeptName); 
\end{codesmall}

\noindent The \texttt{==} 
operation (corresponding to a higher-order function that
constructs a lambda term checking for equality in the output of two input lambda terms) is actually implemented as C++ template
whose two type parameters \texttt{LHSType} and \texttt{RHSType} are inferred from
the output types of the two input lambda terms. 
The \texttt{==} template returns an object of type \texttt{EqualsLambda} \texttt{<LHSType,} \texttt{RHSType>}, which
itself has an operation returning a pointer to the pipeline stage \texttt{Join\_2212.==\_3} referred to in the TCAP.
As expected, this stage repeatedly takes in an input vector list,
creating a new vector of booleans, containing the truth values of the equality check of each \texttt{LHSType} from the left
input vector and each \texttt{RHSType} from the right input vector.
Using C++'s template metaprogramming facilities, this 
pipeline stage is generated specifically for \texttt{LHSType} and \texttt{RHSType} and optimized by the compiler for use with those
two types.  
As the \texttt{Join\_2212.==\_3} pipeline stage loops over the objects in the input vectors, 
there are no function calls that cannot themselves be inlined by the compiler
and optimized---unless, of course, the (potentially) user-defined equality operation over \texttt{LHSType} and \texttt{RHSType}
objects itself contains a virtual function call.

In this way, each pipeline stage in the graph described by a TCAP program is generated using template metaprogramming.
Actually pushing a vector list through a stage requires no per-data-object virtual function calls, and the pipeline stages
are generated specifically for the types pushed through the pipeline.

%% file: object_model.tex
\section{Details of the Object Model} 
\label{sec:ObjectModel}

At the core of the system is the PC object model, which allows 
programmers to create, manipulate, and store persistent objects.
In keeping with our vision of granting programmers fine-grained control over how data are managed in the small, the PC object model
is much lower level than what is found in systems targeted more towards application programming, yet still provides a great deal of
key functionalities.

\subsection{PC Objects}

Arguably, the choice of how individual data items are to be represented and manipulated
in a data analytics or management system is one of the most
controversial decisions
that a system designer can make, both in terms of 
the programmability of the resulting system, and its performance.  For decades, the dominant model used in
data management was the flat relational model, which 
can achieve very good performance.
Flatness generally means 
that there is typically no distinction between the in-memory representation of data, and the on-disk (or in-network) representation of
data. Thus there is no (de-)serialization cost to move data to/from
disk and network, and memory management costs are very low. 

The downside is that flat relations are very limiting to a programmer.  Modern, object-based 
data analytics systems 
(such as Spark via its Resillient Distributed Dataset (RDD) interface \cite{zaharia2012resilient}) offer far more flexibility, at the (possible) cost of significant performance degradation.  
PC attempts to combine this flexibility with excellent performance.
The PC object model provides a fully object-oriented interface, supporting the standard functionality expected in a modern, object-oriented type system,
including generic programming (the PC object model supports generic Map and Vector types), pointers (or, more specifically,
``pointer-like'' objects called PC \texttt{Handle} objects), inheritance, and dynamic dispatch for runtime polymorphism.  

For example, imagine that the goal is implementing a distributed linear algebra system on top of the PC object model, where huge matrices are ``chunked'' into
smaller sub-matrices.  A sub-matrix may be stored via our current, C++ binding, using the following object:

\begin{code}
class MatrixBlock : public Object {
public:
	int chunkRow, chunkColumn;
	int chunkWidth, chunkHeight;
	Vector <double> values; 
};
\end{code}

or, a sparse sub-matrix may be stored as:

\begin{code}
class SparseMatrixBlock : public Object {
public:
	int chunkRow, chunkColumn;
	int chunkWidth, chunkHeight;
	Map <pair <int, int>, double> values; 
};
\end{code}

In the sparse sub-matrix, the \texttt{pair <int, int>} indexes a non-zero entry in the chunk by its row and column.

But while the PC object model provides a rich, object-oriented programming model, it also provides the good performance characteristic
of a flat relational model.
The key principle underlying the PC object model is \emph{zero-cost data movement}.  That is, once a data object
has been allocated and populated, moving the object to disk or across the network should be a simple matter of copying memory; there
should be no CPU cost for serialization and deserialization.

At first glance, it would seem to be impossible to offer zero-cost data movement while allowing a programmer to create and manipulate such objects.  
Pointers and container classes 
generally lead to high memory (de)allocation costs and high object (de)serialization costs, resulting in high CPU cost.
The PC object model avoids this 
by using a ``page-as-a-heap'' memory allocation model.  
The PC object model provides a call of the form:

\begin{code}
makeObjectAllocatorBlock (ptr, blockSize);
\end{code}

After such a call, \emph{all subsequent PC Object allocations by the thread creating the object allocation block will be performed directly to the memory
region starting at location} \texttt{ptr}.
Typically, when it runs a computation, PC's execution engine will obtain a page from its buffer pool to buffer output data, calling
PC's
\texttt{makeObjectAllocatorBlock ()} function with a pointer to the page where output data are to be written.  
When an action taken by the execution engine or user-supplied code causes an
out-of-memory execution, it means that the page is full.  At that point, the execution engine can take appropriate computation-specific 
action, such as creating 
an object allocation block out of a new (empty) page, writing the full page out to disk, sending it across the network, etc.  
No serialization or deserialization or any sort of post-processing of the page are needed, 
because all object allocations have taken place exclusively to the current allocation block.  

In order to guarantee zero-cost data movement,
one rule that a PC programmer must follow is that any object that will
be loaded into a distributed PC cluster must either be of a ``simple'' type (a simple type
must contain no raw C-style pointers and no 
virtual functions, and a \texttt{memmove}
must suffice to copy the object), or else it must descend from PC's \texttt{Object} class, which serves as the base for all complex object types.
Complex objects are those that include containers (\texttt{Vector}, \texttt{Map}) or pointer-like \texttt{Handle} objects.  Descending from PC's \texttt{Object} class
ensures that the resulting class type has a set of virtual functions
that allow it to be manipulated in and transferred across the
distributed PC cluster, 
such as a virtual deep copy function.  

\subsection{PC Handles}

To support linked data structures, dynamic allocation, and runtime polymorphism, it is necessary for a 
system to provide pointer-like functionality.  This is provided by PC's built-in \texttt{Handle} type.  
A \texttt{Handle} to an object is returned from a dynamic allocation to the current allocation block.
For example, a PC programmer 
can issue the statement:

\begin{code}
Handle <MatrixBlock> mySubMatrix = makeObject <MatrixBlock> ();
\end{code}

Internally, PC \texttt{Handle} objects contain two pieces of data: an \emph{offset pointer} that tells how far
the physical address of the object being pointed to is from the physical location of the 
\texttt{Handle}, and a \emph{type code} that stores the type of the
object that is pointed to.  

PC uses an offset pointer rather than a classical, C-style pointer in order to support
zero-cost data movement.  
A \texttt{Handle} may begin its life allocated
to one page, which may be stored on disk, then sent across a network
to another process.  
An actual C-style pointer
cannot survive translation from one process to another, as the program will be mapped to a different location in memory.
In contrast, at the new process, the \texttt{Handle} pointer
can function correctly. As long as the target of an offset pointer is stored in the same page, an offset pointer will be valid if the
page is copied in its entirety, including all \texttt{Handle}s and their targets.

\subsection{Dynamic Dispatch}
\label{sec:dyn_dis}

Supporting dynamic dispatch for virtual function calls is fundamental to the PC object model.
In PC, dynamic dispatch is facilitated by the type code stored within each
\texttt{Handle} object.
Each type code begins with a bit that denotes whether or not the referenced type is a simple type (which, by definition, cannot have any
virtual functions and for which a \texttt{memmove} suffices to perform a copy) or a type descended from PC's \texttt{Object} base class.
In the case of a simple type, the remaining bits encode the size of the referenced object.  

In the case of a PC \texttt{Object} or its descendants, the
type code is a unique identifier for the PC-\texttt{Object}-descended type of the object that the \texttt{Handle} points to.
In every major C++ compiler (GCC, clang, Intel, and Microsoft), virtual functions
are implemented using a virtual function table, or \texttt{vTable} object, a pointer to which is located at the beginning
of each C++ object having a virtual function.  Unfortunately, the \texttt{vTable} pointer is a native, C-style pointer, the
\texttt{vTable} pointer does not automatically translate when an
object is moved from process to process.  To handle this, in
PC's C++ binding, whenever a PC \texttt{Handle} object is dereferenced, 
a lookup on the
type code is performed transparently to the application programmer.  This lookup retrieves a process-specific pointer to that class' \texttt{vTable} object, which is
then placed at the head of the object.

Obtaining a pointer to a class' \texttt{vTable} object is not straightforward.
A user may run code on his/her machine that creates
a PC \texttt{Object}, and then ship that PC \texttt{Object} into the
PC cluster.  At the other end, it arrives at a PC worker process that has never seen that type of object before and hence
does not have access to a \texttt{vTable} pointer for that class.
PC addresses this issue by requiring that all classes deriving from PC's \texttt{Object} base class be registered with the PC catalog
server before they are loaded into the distributed storage subsystem.  This registration requires shipping a library file (a \texttt{.so} file in Linux/Unix) to
the catalog server.  This library exposes a special
\texttt{getVTablePtr ()} function that returns a C-style raw pointer to the \texttt{vTable} for the class contained
in the \texttt{.so} file.

Whenever there is a \texttt{vTable} pointer lookup, the request first goes to the PC process' \texttt{vTable} lookup table.  When this 
lookup fails (because the process has not yet seen a \texttt{vTable} pointer for that class
type) the request then goes to the PC cluster's catalog server, which responds to the process with a copy of the appropriate \texttt{.so} file.  This 
\texttt{.so} file is then 
dynamically loaded into the process' address space, \texttt{getVTablePtr ()} is called, 
and the located \texttt{vTable} pointer is loaded into the lookup table, and then copied into the PC \texttt{Object} that is being referenced.  

In this way, PC provides something akin to the automated,
dynamic loading of classes (via Java Virtual Machine \texttt{.class} files) that is
provided by most big data systems.  
Objects of arbitrary type can be loaded into the distributed PC cluster and be processed using dynamically-loaded native code, as long
as the object type is registered first.

\subsection{Allocation, Deallocation, and Cross-Block Assignment}

There are three types of allocation blocks in PC, where an ``allocation block'' is a block of memory where PC \texttt{Object}s can be
allocated, or where they are located.

\begin{enumerate}

\item Each thread running in a PC process has exactly one \emph{active} allocation block, that is currently receiving allocations (all calls to
\texttt{makeObject} cause memory allocations to happen using that block).  Such an allocation block is created via a call to 
\texttt{makeObjectAllocatorBlock ()}.  User code typically creates and manipulates objects in this block.

\item Each thread also has one or more
\emph{inactive}, \emph{managed} blocks.  These are previously-active blocks of memory that contain one or more objects that are reachable
from some \texttt{Handle} that is currently in RAM.  When the number of reachable objects in an inactive, managed block drops to zero, it is automatically
deallocated.
When a user (or the PC system software) calls 
\texttt{makeObjectAllocatorBl ock ()}, the newly created allocation block becomes the active block, and if the previously-active allocation block has any
reachable objects on it, it becomes an inactive, managed block.

\item Finally, there are zero or more \emph{inactive},
  \emph{un-managed} blocks.  These are blocks with reachable PC
  \texttt{Object}s that are \emph{not} managed by the PC object model.  These tend to be pages of objects that have
been loaded into RAM from disk or across the network for processing
during a distributed computation.  Such blocks are paged in and out of the
buffer pool in much the same way as a relational database would page data in and out.
Rather than the PC object model being responsible for managing such blocks, PC's execution engine manages such blocks.
Further, since managed blocks are only managed by the ``home'' thread where
they are created, a managed block is effectively un-managed when viewed from
any other thread.

\end{enumerate}

In PC, each managed allocation block (active or inactive) has an active object counter (the number of objects that are reachable
from some \texttt{Handle} in RAM).  Each object in each managed allocation block (active or inactive) is reference counted, or pre-pended with a count of
the number of \texttt{Handle} objects that currently reference the object.  
Un-managed blocks (and objects inside of such blocks) are not reference-counted.

When the reference count on an object in a managed block goes to zero, it is automatically
deallocated (at least, this is the default behavior; it is possible
for a programmer to override this behavior for speed, if desired, as
we describe in Section~\ref{sec:allocation} in the Appendix).  
Once the number of reachable objects on an inactive, managed allocation block falls to zero, the block is automatically deallocated.  
In that sense, PC resembles a smart-pointer based memory management system.  

Since the fundamental goal of PC object model design is 
zero-cost data movement---an allocation block should be transferable across processes and immediately usable with no pre- or post-processing---one
potential problem is dangling \texttt{Handle}s.  Specifically: What happens when there is a \texttt{Handle} located in one allocation block that points to a PC
\texttt{Object} located in another allocation block?  The \texttt{Handle} may be valid, but when the \texttt{Handle}'s allocation block is moved to a new process where
the target block is not located, the \texttt{Handle} cannot be dereferenced without a runtime error. 
PC simply prevents this situation from ever happening. Whenever an assignment operation on \texttt{Handle} that is physically located
in the active allocation block results in that
\texttt{Handle} that is physically located in the active allocation block pointing outside of the block, a deep copy of the target of the assignment
is automatically performed.  This deep copy happens recursively, so any \texttt{Handle}s in the copied object that point outside of the active allocation block
have \emph{their} targets deep-copied to the active block.  For example, consider the following code:

\begin{code}
makeObjectAllocatorBlock (1024 * 1024);
Handle <Vector <double>> data = makeObject <Vector <double>> ();
for (int i = 0; i < 1000; i++)
     data->push_back (i * 1.0);
makeObjectAllocatorBlock (1024 * 1024); 
Handle <MatrixBlock> myMatrix = makeObject <MatrixBlock> ();
myMatrix->value = data; // deep copy of data happens!!
\end{code}

At the second \texttt{makeObjectAllocatorBlock}, the original allocation block, holding the list of \texttt{double}s pointed to by \texttt{data}, becomes
inactive.  The submatrix \texttt{myMatrix} is allocated to the new active block.  
Hence, the assignment of \texttt{data} to \texttt{myMatrix->value} is cross-allocation block, and a deep copy automatically happens to ensure that
the current block is zero-cost copy-able and movable.  

Such cross-block assignments require deep copies and are expensive,
but in practice, such they are rare, and a programmer who understands the cost can often avoid them, making sure to allocate
data that must be kept together to the same block.
Again, this is in-keeping with PC's design philosophy: trust the ability of the programmer to do the right thing, in the small.

\subsection{The PC Object Model and Multiple Threads}

While smart-pointer-based memory management
systems are often significantly faster than garbage collected systems, such systems still have their bottlenecks.  One of the bottlenecks is concurrency
control.
Since an object can have pointers across multiple threads, smart pointer counters must be locked before increment/decrement, which
can have a significant impact on performance.  PC, however, does not need to lock reference counts (or active object counts) because only managed blocks maintain
object reference counts and active object counts,
and a block can only be managed by a single thread.  
If a thread copies a \texttt{Handle} object referencing an object housed on another thread's managed block, 
the reference count will not be changed because from the copying thread's point-of-view, the allocation block is not managed.
This can, in theory, result in a problematic case where one thread has a \texttt{Handle} to an object that has been deallocated on the other thread (since
the reference count on the home thread will not be updated to reflect
the off-thread reference).  But in practice, it tends not to be a
problem.  Parallel and distributed processing is transparent to PC application
programmers, and they typically do not write explicitly multi-threaded code, so most cross-thread references happen as the result of computations staged by the 
PC execution engine.  The PC execution engine typically uses pages carefully so as to ensure that 
it is not possible for pages to be unpinned while references to them can still exist.

%% file: tcap_optimization.tex
\section{Optimizing TCAP}
\label{sec:optimizer}

One of the key ideas driving the design and implementation of PlinyCompute is that \emph{all} PC computations should be optimized, both to
match programmer expectation---programmers generally expect
that changes in the way that boolean expressions are composed should not affect system runtime---and 
to protect against poor programmer choices when constructing the query graph.

Optimizability is one of the drivers for
the decision to compile all computations expressed in PC's lambda calculus into TCAP.  
TCAP resembles relational algebra, and it is similarly amenable to rule- and cost-based optimization
using a combination of methods from relational query optimization and classical compiler construction.
Currently, the optimizations implemented in PC are rule-based (such as pushing down selections).  We plan to work on cost-based optimization
in the future---this is a challenging research problem because of a lack of statistics over the data, which are arbitrary PC \texttt{Object}s.

PC's optimizer is currently implemented
in Prolog; a series of transformations are fired iteratively to improve the plan until the plan cannot be improved further.
For an example of the sort of optimization present in PC, consider the task of removing redundant method calls.  Imagine that a user
supplies a \texttt{SelectionComp} with the following \texttt{getSelection ()}:

\begin{codesmall} 
Lambda <bool> getSelection (Handle <Emp> emp) {
        return makeLambdaFromMethod (emp, getSalary) > 50000 &&
		makeLambdaFromMethod (emp, getSalary) < 10000;
}	
\end{codesmall}

\noindent PC would compile this into the following TCAP:

\begin{codesmall}
JK2_1(emp,mt1) <= APPLY(In(emp), In(emp), 'Sel_43', 'method_call_1',
  [('type', 'methodCall'), ('methodName', 'getSalary')]);

JK2_2(emp,bl1) <= APPLY(JK2_1(mt1), JK2_1(emp), 'Sel_43', '>_1', 
  [('type', 'const_comparison'), ('op', '>')]);

JK2_3(emp,bl1,mt2) <= APPLY(JK2_2(emp), JK2_2(emp,bl1), 'Sel_v3', 'method_call_2',
  [('type', 'methodCall'), ('methodName', 'getSalary')]);

JK2_4(emp,bl1,bl2) <= APPLY(JK2_3(mt2), JK2_3(emp,bl1), 'Sel_43', '<_1',
  [('type', 'const comparison'), ('op', '<')]);

JK2_5(emp,bl3) <= APPLY(JK2_4(bl1,bl2), JK2_4(emp), 'Sel_43', '&&_1', 
  [('type', 'bool_and')]);

JK2_6(emp) <= FILTER(JK2_5(bl3), JK2_5(emp), 'Sel_43', []);
\end{codesmall}

\noindent
This TCAP program first calls the method \texttt{getSalary ()} on \texttt{In::emp} to produce a new vector list \texttt{JK2\_2}, storing the result
of the method call in \texttt{JK2\_1.mt1}.  After comparing \texttt{JK2\_2.bl1} to \texttt{50000}, the result of the method call is dropped.
The method is then called once again on \texttt{JK2\_2.emp} and the result compared with \texttt{100000} to produce \texttt{JK2\_4}, at which 
point the two boolean vectors are ``anded'' and the result is filtered.

Obviously, there is a redundancy here as the method \texttt{getSalary ()} will be called twice.
If \texttt{getSalary} simply accesses a data member, the additional
call is costless.  But in the general case, a method call may run an arbitrary
computation.  Hence, the second call should automatically be removed as being redundant 
(by definition, all method calls evaluated during computation should
be purely functional, and so they must return the same value when called a second time).
The TCAP optimization rule leading to its removal is:

\begin{itemize}
\vspace{-5 pt}
\item If two \texttt{APPLY} operations are both of type \texttt{methodCall} and both invoke the same \texttt{methodName};
\vspace{-5 pt}
\item And one \texttt{APPLY} operation is the ancestor of the other in the TCAP graph;
\vspace{-5 pt}
\item And both \texttt{APPLY} operations operate over the same data object;
\vspace{-5 pt}
\item Then the second \texttt{APPLY} operation can be removed, and the result of the first \texttt{APPLY} carried through the graph.
\end{itemize}

\noindent
In our example, the
optimized
TCAP program is:

\begin{codesmall}
JK2_1(emp,mt1) <= APPLY(In(emp), In(emp), 'Sel_43', 'method_call_1',
   [('type', 'methodCall'), ('methodName', 'getSalary')]);

JK2_2(emp,mt1,bl1) <= APPLY(JK2_1(mt1), JK2_1(emp,mt1), 'Sel_43', '>_1', 
  [('type', 'const comparison'), ('op', '>')]);

JK2_4(emp,bl1,bl2) <= APPLY(JK2_3(mt1), JK2_3(emp,bl1), 'Sel_43', '<_1', 
  [('type', 'const comparison'), ('op', '<')]);

JK2_5(emp,bl3) <= APPLY(JK2_4(bl1,bl2), JK2_4(emp), 'Sel_43', '&&_1', 
  [('type', 'bool_and')]);

JK2_6(emp) <= FILTER(JK2_5(bl3), JK2_5(emp), 'Sel_43', []);
  
\end{codesmall}

\noindent
For another example of a rule-based TCAP optimization, 
consider the classical technique of pushing selection predicates past joins.  Imagine that a user supplied the following
\texttt{getSelection ()} for a \texttt{JoinComp} operation:

\begin{codesmall} 
Lambda <bool> getSelection (Handle <Emp> emp, Handle <Emp> sup) {
        return makeLambdaFromMethod (emp, getSalary) > 50000 &&
		(makeLambdaFromMethod (emp, getSupervisor) == 
                 makeLambdaFromMember (sup, name));
}	
\end{codesmall}

\noindent
Since all selection predicates are by default evaluated \emph{after} the join, 
this would be compiled to the following TCAP code:

\begin{codesmall}
JK2_1(sup,mt1) <= APPLY(InSup(sup), InSup(sup), 'Join_42', 'att_access_1', 
  [('type', 'attAccess'), ('attName', 'name')]);

JK2_2(sup,hash1) <= HASH(JK2_1(mt1), JK2_1(sup), 'Join_42', []);

JK2_3(emp,mt2) <= APPLY(InEmp(emp), InEmp(emp), 'Join_42', 'method_call_1', 
  [('type', 'methodCall'), ('methodName', 'getSupervisor')]);

JK2_4(emp,hash2) <= HASH(JK2_3(mt2), JK2_3(emp), 'Join_42', []);

JK2_5(sup,emp) <= JOIN(JK2_2(hash1), JK2_2(sup), 
  JK2_4(hash2), JK2_4(emp), 'Join_42', []);

JK2_6(sup,emp,mt3) <= APPLY(JK2_5(emp), JK2_5(sup,emp), 'Join_42', 'method_call_2',
  [('type', 'methodCall'), ('methodName', 'getSalary')]);

JK2_7(sup,emp,bool1) <= APPLY(JK2_6(mt2), JK2_7(sup,emp), 'Join_42', '>_1', 
  [('type', 'const comparison'), ('op', '>')]);

/* additional code here to check whether getSupervisor == name... 
   result goes into JK2_10.bool2 */

JK2_11(sup,emp,bool3) <= APPLY(JK2_10(bl1,bl2), JK2_10(sup,emp), 'Join_42', '&&_1', 
   [('type', 'bool_and')]);

JK2_12(sup,emp) <= FILTER(JK2_11(bool2), JK2_11(sup,emp), 'Join_42', []);
\end{codesmall}
 
\noindent
This code first uses \texttt{emp.getSupervisor ()} and \texttt{sup.name} to obtain the join keys. These are hashed, and 
a hash join is run (this is the \texttt{JOIN} operation).  After the hash join,
the result of calling \texttt{emp.getSuper visor ()} is compared with
\texttt{sup.name}.  If these two values are equal and the salary exceeds \texttt{50000}, the result tuple is accepted.

Clearly, it should be possible to first filter based off of the salary exceeding \texttt{50000} before the hash join is ever run.  Hence, one of
the rule-based optimizations available to PC is that:

\begin{itemize}

\vspace{-5 pt}
\item If there is a boolean predicate of the form $(b_1 \wedge b_2 \wedge ...)$ that operations
on the result of a join;

\vspace{-5 pt}
\item And some $b_i$ refers to values that depend only on one of the join inputs (in this case, \texttt{emp.getSupervisor ()}
\texttt{>} \texttt{50000};
depends only upon \texttt{emp});

\vspace{-5 pt}
\item Then $b_i$ can be pushed down to that join input, and a new
  \texttt{FILTER} is introduced.
\end{itemize}

\noindent In this case, after the transformation, we would have:

\begin{codesmall}
JK2_1(sup,mt1) <= APPLY(InSup(sup), InSup(sup), 'Join_42', 'att_access_1', 
  [('type', 'attAccess'), ('attName', 'name')]);

JK2_2(sup,hash1) <= HASH(JK2_1(mt1), JK2_1(sup), 'Join_42', []);

JK2_6(emp,mt3) <= APPLY(InEmp(emp), InEmp(emp), 'Join_42', 'method_call_2', 
  [('type', 'methodCall'), ('methodName', 'getSalary')]);

JK2_7(emp,bool1) <= APPLY(JK2_6(mt2), JK2_7(emp), 'Join_42', '>_1', 
  [('type', 'const comparison'), ('op', '>')]);

JK_2_7_1(emp) <= FILTER(JK2_7(bool1), JK2_7(emp), 'Join_42', []);

JK2_3(emp,mt2) <= APPLY(JK_2_7_1(emp), JK_2_7_1(emp), 'Join_42', 'method_call_1', 
  [('type', 'methodCall'), ('methodName', 'getSupervisor')]);

JK2_4(emp,hash2) <= HASH(JK2_3(mt2), JK2_3(emp), 'Join_42', []);

JK2_5(sup,emp) <= JOIN(JK2_2(hash1), JK2_2(sup), JK2_4(hash2), 
  JK2_4(emp), 'Join_42', []);

/* additional code here to check whether getSupervisor == name... 
   result goes into JK2_10.bool2 */

JK2_11(sup,emp,bool3) <= APPLY(JK2_10(bl1,bl2), JK2_10(sup,emp), 'Join_42', '&&_1', 
  [('type', 'bool_and')]);

JK2_12(sup,emp) <= FILTER(JK2_11(bool2), JK2_11(sup,emp), 'Join_42', []);
\end{codesmall}

%% file: experiments.tex
\section{Experiments}
\label{sec:exp}
\subsection{Overview}

In this section, we describe our experimental evaluation of PC.
The aim is to answer
the following questions:

\begin {enumerate}
\item PC has been designed to facilitate the construction of high-performance Big Data tools and libraries by programmer
comfortable with lower-level systems concepts, including memory management.
How useful is PC for this task?
\item PC's core design goals were ``declarative in the large,
high-performance in the small.'' The latter goal---high performance in the small---was largely realized via the 
PC object model.  We wish to answer the question: Can the PC object model be used to build 
object-oriented computations
that efficiently manipulate highly nested and complex objects?
\item Finally, machine learning (ML) is an important component of many libraries---and ML will only increase in importance
as a target for tool and library development in the future.
So we wish to ask: How well does PC compare to alternative systems for developing scalable ML algorithm implementations?
\end {enumerate}

\noindent In an attempt to answer each of these questions, we perform three different benchmarking tasks:

\begin {enumerate}
\item To test the applicability of PC for tool and library construction, we constructed a scalable, distributed
linear algebra library
called \texttt{lilLinAlg} on top of PC, and evaluated \texttt{lilLinAlg}'s performance for running three
computations that could reasonably be expressed in linear algebra: distributed Gram matrix construction, 
distributed least squares linear regression, and distributed nearest neighbor search.

\item To test the utility of the PC object model, we first denormalized the TPC-H database \cite{council2008tpc} into an object-oriented 
representation, and then benchmarked two reasonably complex 
analytical computations ---the
first computes the list of customers and the parts they construct for each supplier, and the
second is a top-$k$ similarity query that searches for the customers whose set of purchased items is most similar
to a query set.

\item Finally, we also implemented three widely used
  iterative machine learning algorithms on top of PC: Latent Dirichlet Allocation (LDA) which is used for
  textual topic mining;
  Gaussian mixture model (GMM) learning, which is used to cluster data using a mixture of high-dimensional Normal
  distributions, and the simplest, $k$-means clustering (chosen because of its ubiquity as a Big Data ML benchmark).
\end {enumerate}

\vspace{5 pt}
\noindent
\textbf{Organization.}  The remainder of this section is organized as follows.  First, we describe the experimental
environment used to evaluate these implementations.  
Next we describe each of the three benchmarks, in sequence. We then conclude the experiments with a general
discussion of the results.  This includes a discussion of implementation complexity, as well as a discussion on how much
our results are (or are not) simply a direct result of coding using C++, as opposed to a managed language such as Java or
C\#.

\subsection {Experimental Environment}

All of the experiments reported in this paper were performed using a
cluster that consists of eleven Amazon EC2 \texttt{m2.4xlarge}
machines. Each machine
ran Ubuntu 16.04, except the linear algebra experiments on
SciDB~\cite{brown2010overview, stonebraker2011architecture}
with version 14.8, which
was supported on Ubuntu 12.04~\cite{SciDBOS}.
Each machine had eight virtual cores, one SSD
disk, and 68 GB of RAM. In each PC cluster that we built, one of the eleven machines served as the master
node and the rest ten machines served as worker nodes.

Since Apache Spark is one of the most widely-used Big Data system both for applications programming and for tool and library
development, most
(though not all) of our comparisons were with Spark (version
2.1.0). For TPC-H and LDA, of which total volume of data for input and
processing exceeded available memory, we ran Spark in yarn client
mode to avoid the out-of-memory errors. For other experiments, we ran Spark
in cluster mode to be consistent with PC.

The configuration of the Spark cluster such as number of executors,
executor memory, number of cores for each executor, driver memory and so on are carefully tuned for each
experiment, as shown in Table~\ref{fig:sparkConf}. We do not clear the OS buffer cache, so HDFS data can be buffered or
cached in the OS buffer cache. 

In addition, input data for experiments
using Dataset APIs were stored in Parquet format, and input data for
experiments using RDD APIs were stored in Spark's object file format,
and serialized using Kryo. Other Spark parameters such as parallelism,
partition number, and so on were all carefully tuned for each experiment. More details
are omitted due to space limitation.

\begin{table}[H]
\begin{center}
\begin{tabular}{|c||c|c|c|c|}
\hline
Platform & \makecell{num executors} & \makecell{executor mem} & \makecell{executor cores}& \makecell{driver mem}\\
\hline
\texttt{lilLinAlg} &10 & 60GB & 8 & 50GB \\
TPC-H &10 & 50GB & 7 & 50GB \\
LDA &20 & 26.5GB$*$ &4 & 55GB\\
GMM&80 & 70GB & 1 & 55GB\\
$k$-means &10 &60GB & 8 & 50GB\\
\hline
\end{tabular}
\caption{Workload-specific Spark Configurations for Different
  Experiments. A star ($*$) indicates additional 4GB off heap memory is used.}
\label{fig:sparkConf}
\end{center}
\end{table}

\subsection {Distributed Linear Algebra}

Since PC is designed to support the construction
of high-performance tools and libraries, our first benchmarking effort was aimed at determining 
whether PC is actually useful for that task.  Thus, we asked
a PhD student (who was expert programer but at the outset knew nothing of PC) 
to use the system to build a small Matlab-like 
programming language and library for distributed matrix operations.
We called this implementation \texttt{lilLinAlg}.

Our goal was to determine the 
performance and functionality that an expert programmer (but PC novice) could deliver in a short
time-frame, compared to a set of established distributed Big Data linear algebra implementations:
SciDB \cite{brown2010overview, stonebraker2011architecture} (built from the ground up by a team
consisting of MIT students and professional developers over the last
nine years), Spark \texttt{mllib} \cite{meng2016mllib} 
(the Big Data matrix
implementation shipped with Spark), and SystemML \cite{boehm2014hybrid, ghoting2011systemml, boehm2016systemml}
(a matrix and machine learning implementation developed
over the last seven years by a team at IBM, built on top of Spark and Hadoop).
The student spent about six weeks in this effort.

\subsubsection{\texttt{lilLinAlg} Implementation}

In \texttt{lilLinAlg}, a distributed matrix is stored as a set of PC \texttt{Object}s, where each 
object in the set is a \texttt{MatrixBlock}, similar with the
\texttt{MatrixBlock} class described in
Section~\ref{sec:ObjectModel}, storing a contiguous rectangular sub-block of the matrix.

%
%
%
%

The actual data stored in a \texttt{MatrixBlock} object should be small enough to fit completely in 
a PC page (by default, PC's page size is 256MB).  A typical \texttt{MatrixBlock} object stores a
1,000 by 1,000 sub-matrix that is eight megabytes in size.

\texttt{lilLinAlg} uses the \texttt{MatrixBlock} object to implement a set of common distributed matrix
computations, including \texttt{transpose},
\texttt{inverse}, \texttt{add}, \texttt{subtract}, \texttt{multiply}, \texttt{transposeMultiply}, 
\texttt{scaleMultiply}, \texttt{minElement},
\texttt{maxElement}, \texttt{rowSum}, \texttt{columnSum},
\texttt{duplicateRow}, \texttt{duplicat eCol}, 
and many more.  However, \texttt{lilLinAlg} programmers do not call these operations directly, rather,
\texttt{lilLinAlg} implements its own Matlab-like DSL.  
Given a computation in the DSL, \texttt{lilLinAlg} first parses the computation into an abstract syntax tree (AST), and then
uses the AST to build up a graph of PC \texttt{Computation} objects which is used to implement the distributed computation.
For example, at a \texttt{multiply} node in the compiled AST, \texttt{lilLinAlg} 
will execute a PC code similar to the following:

\begin{codesmall}
Handle <Computation> query1 = makeObject <LAMultiplyJoin> ();
query1->setInput (0, leftChild->evaluate(instance));
query1->setInput (1, rightChild->evaluate(instance));
Handle <Computation> query2 = makeObject <LAMultiplyAggregate> ();
query2->setInput(query1);
\end{codesmall}

\noindent Here, \texttt{LAMultiplyJoin} and \texttt{LAMultiplyAggregate} are 
both user-defined \texttt{Computation} classes that are
derived from PC's \texttt{JoinComp} class and \texttt{AggregateComp} class, respectively; these classes are chosen 
because distributed matrix multiplication
is basically a join followed by an aggregation.
Internally, the \texttt{LAMultiplyJoin}
and \texttt{LAMultiplyAggregate} invoke the Eigen numerical processing
library~\cite{eigen} to manipulate \texttt{MatrixBlock} objects.
For example, \texttt{LAMultiplyJoin} must efficiently multiply the sub-matrices stored inside of two matrix blocks.
To accomplish this, inside of the \texttt{getProjection ()} operation for
\texttt{LAMultiplyJoin} is a native C++ lambda that contains the following code:

\begin{code}
Handle<MatrixBlock> resultMatrixBlock = makeObject <MatrixBlock> (...);
Eigen::Map<Eigen::Matrix<double,Eigen::Dynamic,Eigen::Dynamic,Eigen::RowMajor>> 
   productMatrix(resultMatrixBlock->getRawDataHandle()->c_ptr(), ...);
productMatrix = currentMatrix1 * currentMatrix2;
\end{code}

\noindent This code first makes a new \texttt{MatrixBlock} object that will store the result of the multiplication.
 It then creates an 
Eigen row major matrix object whose storage is located at 
\texttt{resultMatrixBlock->}\-\texttt{getRaw}\-\texttt{DataHandle()->}\-\texttt{c\_ptr()}.  This
sequence of calls
returns a \texttt{double} pointer that references the location of the contiguous block of \texttt{double}s inside of
the \texttt{resultMatrixBlock} object.
The Eigen package then works with those raw bytes directly,
and the overloaded \texttt{*} operator will write its result to that location.

\texttt{lilLinAlg}'s DSL looks a lot like Matlab and allows very short and easy-to-read codes.
For example, 
a least squares linear regression over a large input matrix can be easily coded as

\begin{code}
X = load(myMatrix.data); 
y = load(myResponses.data); 
beta = (X '* X)^-1 
\end{code}

\noindent In the above DSL expression, \texttt{'*} represents a transpose-then-multiply computation,
\texttt{\^{}-1} represents an inverse computation, and \texttt{ \%*\%}
represents a multiply computation. 

\subsubsection {Experiments}

Our experimental benchmark consisted of three different computations:
a Gram matrix computation (given a matrix $\textbf{X}$, compute
$\textbf{X}^T \textbf{X}$), least squares linear regression (given a matrix of features $\textbf{X}$ and
responses $\textbf{y}$, compute 
$\hat{\pmb{\beta}} = (\textbf{X}^{T} \textbf{X})^{-1} \textbf{X}^{T} \textbf{y}$), and nearest
neighbor search in a Riemannian metric space \cite{lebanon2006metric} encoded by matrix $\textbf{A}$ (that is,
given a query vector
$\textbf{x}'$ and matrix $\textbf{X}$, find the $i$-th row in the matrix that minimizes 
$d_{\textbf{A}}^2(\textbf{x}_i, \textbf{x}') = 
(\textbf{x}_i - \textbf{x}')^T\textbf{A}(\textbf{x}_i - \textbf{x}')$).  
For each computation we used three different data dimensionalities: ten, $10^2$, and $10^3$.  This refers to
the number of features or entries in each data point. 
For
all three computations, 
$10^6$ data points were used. 

In addition to \texttt{lilLinAlg}, 
for the Gram matrix and linear regression computations, SystemML V0.9 on Hadoop was used.
For these two computations, Spark \texttt{mllib} along with
Spark 1.6.1 was used.  For
nearest neighbor, SystemML V1.0 on Spark 2.1.0 was used, and for
nearest neighbor, \texttt{mllib} along with Spark 2.1.0 was used. We
use the same SciDB version---14.8---for all three
experiments.

We spent considerable effort tuning all of the implementations.
For \texttt{lilLinAlg}, this consisted mainly of efforts to choose the correct page size for holding the
\texttt{MatrixBlock} objects.  The task was balancing the ability to fully distribute the computations
(which requires a large number of small \texttt{MatrixBlock} objects) versus making sure that the computations
themselves were efficient (which requires large \texttt{MatrixBlock} objects).  We settled on 
a 4 MB page size for ten dimensions, a 16 MB page
size for $10^2$ dimensions, and a 64 MB page size for $10^3$ dimensions. 
PC's query optimizer dynamically decided to use a
broadcast join to implement matrix operations when one input to the join is smaller than 
two
gigabytes.  Otherwise, it uses a full hash partition join.

For the runs on the other three platforms, we also carefully tuned the
systems for best performance. For example, we tuned Spark block size and repartition
size for every experiment. In SystemML, we also carefully chose to
use the parallel for loop (\texttt{parfor}), which boosted performance significantly.

For fairness, for each of the distributed linear algebra tools, 
we do not count the time required to load data from the client into the system
(for example, 
for \texttt{lilLinAlg}, we do not count the time required to load data from text and into PC).

\subsubsection {Results and Discussion}

Experimental results are given in 
Table \ref{fig:LR}. 
This table shows that for every one of the higher-dimensional computations, the 
\texttt{lilLinAlg} implementation was the fastest.  Often, it was considerably faster.
Looking only at the nearest neighbor computation (where the latest version of Spark was used
along with Spark's \texttt{mllib}) \texttt{lilLinAlg} was five times faster than \texttt{mllib}
and thirteen times faster than SciDB.  

For the smallest, ten-dimensional computations, there was some variability in the results.
For two of the three computations, 
SystemML was the fastest.  However, in all three of the ten-dimensional computations, SystemML chose
\emph{not} to distribute the underlying computation, as it was small enough to be efficiently extracted
on a single machine.  This demonstrates that for a small computation, the overhead of performing it in 
distributed fashion across multiple machines calls into question the viability of distribution in the first place.

\begin{table}[h!]
\begin{center}
\begin{tabular}{|c||c|c|c||c|c|c||c|c|c|}
\hline
& \multicolumn{3}{c||}{Gram Matrix} & \multicolumn{3}{c||}{Linear Regression} & \multicolumn{3}{c|}{Nearest Neighbor} \\
\hline
Dimensionality & $10$ & $100$ & $1000$& $10$ & $100$ & $1000$& $10$ & $100$ & $1000$ \\
\hline
\hline
PC (\texttt{lilLinAlg}) &00:07 & 00:09 &00:39 &00:14 &00:22 &00:49& 00:15 & 00:20 & 01:06 \\
SystemML &00:05$*$ &00:51 &02:34 &00:06$*$ &00:53 &02:38 &00:04$*$ &00:30 &01:32 \\
Spark \texttt{mllib} &00:20  &00:54 &17:31 &00:35 &01:01 &17:42 &01:20 & 04:49 &14:30 \\
SciDB   &00:03 &00:17 &03:20 &00:15 &00:33 &06:04 &00:28 &02:56 & 06:24 \\
\hline
\end{tabular}
\caption{Linear algebra benchmark. Format is MM:SS.
A star ($*$) indicates running in local mode.}
\label{fig:LR}
\end{center}
\end{table}

We feel that overall, these results largely validate the hypothesis that PC is an excellent platform for the 
construction of Big Data tools and libraries.  The only distributed linear algebra implementation
to approach \texttt{lilLinAlg}s performance on the larger matrices
was SystemML.  The newest SystemML version, on Spark, is only 50\% slower than \texttt{lilLinAlg} for nearest neighbor search.
However,
SystemML was built over many years by a team of PhDs, and research papers have been written about the
technology developed for the system, including one awarded a VLDB best paper award \cite{boehm2016systemml}.
\texttt{lilLinAlg} was developed in six weeks by a single PhD student, and it is still
faster (though to be fair, SystemML has a much broader
set of capabilities than \texttt{lilLinAlg}).
One may conjecture that had SystemML been built on a platform such as PC rather than on Spark, it might be significantly
faster than it is now.

Despite the demonstrated benefits of building \texttt{lilLinAlg} on top of PC, we point out that
PC is a young system and so it is still missing some key functionality that would boost
\texttt{lilLinAlg}'s performance even more.  For example, PC cannot make use of pre-partitioning of the data stored in a set. 
If the
\texttt{MatrixBlock} objects making up a distributed
matrix could be pre-partitioned based upon the row/column at load time, it 
would mean that the expensive join for an operation
such as multiply could completely avoid a runtime partitioning of the data, which requires shuffling each input matrix.
Thus, it is not unreasonable to suggest that as PC matures, it will be even faster.

\subsection{Big Object-Oriented Data}

Programming with objects is attractive as a programming paradigm, but (as we have argued in this paper)
often costly in terms of performance, particularly
for distributed computing.  One answer is to simply disallow complex objects.  The 
developers of Apache Spark, for example, have attempted to move away from object programming
and towards a relational model of programming (with Datasets and Dataframes) 

Our solution is to allow objects, but to move away from allowing a managed environment to control 
issues such as allocation, deallocation, and data placement.  This is the approach taken in the design and implementation
of the PC object model.  Thus, the question we address in this particular set of experiments is: can the PC 
object model facilitate efficient computations of heavily nested, complex objects?

\subsubsection{Data Representation}

To do this, we implement two different complex object computations on top of PC and on top of Spark.
Both of these computations are over large datasets that store an instance of the TPC-H database \cite{council2008tpc}.
But rather than storing the dataset relationally, we denormalized the data into a set of nested objects. 
The simplest objects used in the denormalized TPC-H schema are 
\texttt{Part} objects, which do not look very different from the records in the  \texttt{part} schema of the TPC-H database.
In PC, these are defined as:

\begin{codesmall}
class Part : public Object {
private:
   int partID;
   String name;
   String mfgr;
   /* six more members... */};
\end{codesmall}

\noindent
\texttt{Supplier} objects are defined similarly.
\texttt{Lineitem} objects contain nested \texttt{Part} and \texttt{Supplier} objects:

\begin{codesmall}
class Lineitem : public Object {
private:
   Supplier supplier;
   Part part;
   int orderKey;
   int lineNumber;
   /* twelve more members... */}; 
\end{codesmall}
 
\noindent Then,
\texttt{Order} objects have a nested list of \texttt{Lineitem} objects, and \texttt{Customer} objects have a nested
list of all of the
\texttt{Lineitem} objects for a given customer:

\begin{codesmall}
class Order : public Object {
   Vector <Handle <LineItem>> lineItems;
   int orderkey;
   int custkey;
   /* seven more members... */};

class Customer : public Object {
   Vector <Handle <Order>> orders;
   int custkey;
   String name;
   /* seven more members... */};
\end{codesmall}

\subsubsection{Implementation and Experiments}
We then run two computations over the resulting set of \texttt{Customer} objects.  
The first computation is the \emph{customers per supplier} computation where we
compute, for each supplier,
the complete list of \texttt{partID}s that the supplier has sold to each of the supplier's customer's.
For each supplier, the result is an object that contains the supplier's name (as a \texttt{String}) and
an object of type \texttt{Map <String, Handle <Vector <int}\texttt{>}\texttt{>}\texttt{>}.  In this
object, the \texttt{String} is the name of the customer, and the \texttt{Vector} stores the list of
\texttt{partID}s sold to that customer.

To run this computation on PC, we use two different PC \texttt{Computation} classes.  The first, 
\texttt{Customer}\-\texttt{MultiSelection}, transforms each 
\texttt{Customer} object to one or more
\texttt{SupplierInfo} objects. Each \texttt{SupplierInfo} contains
the name of a supplier
and a \texttt{Handle} to a \texttt{Map} whose key is the name of the customer and whose value is the list
of \texttt{partID}s that the supplier has sold to the customer.
The second \texttt{Computation}, called
\texttt{CustomerSupplierPartGroupBy},
groups all of those \texttt{SupplierInfo} objects according to the name of the supplier, computing, for each
supplier, the map from customer name to \texttt{Vector} of \texttt{partID}s.

On Spark we implemented an algorithmically equivalent
carefully-tuned code that in the
end was not dis-similar from the PC
version.  We used
Spark version
2.1.0.
Note that since the objects are all highly nested, it was not possible to develop a satisfactory
Dataset or Dataframe implementation, and so our Spark implementation made use of Spark's RDD interface.  All
implementation was done in Java.  Since Spark makes use of lazy evaluation, it is not possible to collect a timing
for this computation unless we actually \emph{do} something with the result.  So in both the PC and the Spark computations
we add a final count of the number of customers in each \texttt{Map} in each \texttt{SupplierInfo} object.

The second computation run over the denormalizd TPC-H schema is the \emph{top-k closest customer part sets} computation.
In this computation, 
for a given
\texttt{Customer} object, we go through all of the associated 
\texttt{Order} objects and obtain the complete list of
\texttt{partID}s for each order.  All duplicate \texttt{partID}s are
removed from this list, and then the Jaccard similarity between the resulting \texttt{partID} list and a special, query
list are computed.  This is done for all of the \texttt{Customer} objects, and the $k$ \texttt{partID} lists with the 
closest similarity to the query list are returned.
In PC,
the result of the computation is a list of $k$ objects containing the Jaccard similarity, the integer \texttt{custkey}, and
a \texttt{Vector <int>} that stores the complete list of unique \texttt{partID}s sold to that customer.
In PC, the C++ code required to drive this computation is as follows:
\begin{code}
Handle<Vector<int>> listOfPartIDs = 
    makeObject<Vector<int>>(199, 22, 34, 567, 1200, 37, 46, 459, ...);
Handle<Computation> myReader = 
    makeObject<ObjectReader<Customer>>("TPCH_db", "tpch_bench_set1");
Handle<Computation> myTopK = makeObject<TopJaccard>(k, *listOfPartIDs);
myTopK->setInput(myReader);
Handle<Computation> myWriter = makeObject<Writer
    <TopKQueue<double, Handle<AllParts>>>>("TPCH_db", "result");
myWriter->setInput(myTopK);
pcContext.executeComputations (myWriter);
\end{code}

\noindent The one query-specific \texttt{Computation} object that was implemented for the top-$k$ 
closest customer part sets computation is the \texttt{TopJaccard} class, which
is responsible for extracting a value to drive the top-$k$ computation (in this case, the Jaccard similarity) as well
as the object to be associated with that value (in this case, the \texttt{custkey} and the list of 
\texttt{partID}s sold to that customer).

For both PC and for Spark, and for both computations, we created 
TPC-H dataset of various sizes: 2.4 million,
4.8 million, 9.6 million, 14.4 million, 19.2 million and 24 million
\texttt{Customer} Objects respectively.  For the ``top-$k$
closest customer part sets'' computation, $k$ was chosen to be $\frac{1024}{2.4 \times 10^6}$ times the size
of the data.

For both Spark computations, we performed two runs at each dataset size.
For the first, we stored the data in HDFS, and measured the time to execute the query starting with a
read from HDFS.
For the second, we made sure that the data were de-serialized and stored in RAM by Spark.
To do this, we applied a \texttt{distinct().count()} operation to
an RDD storing \texttt{Customer} objects (thus ensuring full deserialization) before running each query.
All data were
serialized using Kryo, and parameters such as data partition size and parallelism are fully tuned to obtain
optimal performance. 

For PC, we ran only one version of the computation, where the various \texttt{Customer} objects are stored in
PC's storage system.  Since all datasets are small enough to be cached in RAM, there is no I/O time to
retrieve data.

\subsubsection{Results and Discussion}

\begin{table}[h!]
\begin{center}
\begin{tabular}{|c||c|c|c|c|c|c|}
\hline
Number \texttt{Customer} objects &2.4M & 4.8M & 9.6M  & 14.4M & 19.2M & 24M \\
Kryo data size &41.5GB & 83.1GB & 167.2GB &251.1GB &333.2 &416.2GB \\
\hline
& \multicolumn{6}{c|}{\texttt{Customer}s per \texttt{Supplier}} \\
\hline
PlinyCompute: hot storage & 00:11&	00:19&	00:35&	00:51&	01:08&	01:21 \\
Spark: hot HDFS & 01:04&	01:53&	03:24&	04:54&	06:25&	08:16\\
Spark: in-RAM deserialized RDD & 00:16& 	00:29& 	00:56& 	01:21& 	02:18& 	03:56\\
\hline
& \multicolumn{6}{c|}{top-$k$ Jaccard} \\
\hline
PlinyCompute: hot storage & 00:03&	00:03&	00:04&	00:05&	00:05&	00:06 \\
Spark: hot HDFS & 00:56&	01:38&	03:01 & 04:01&	05:22&	06:34\\
Spark: in-RAM deserialized RDD & 00:08& 	00:12& 	00:21 & 00:32& 	01:11& 	02:38\\
\hline
\end{tabular}
\caption{PlinyCompute vs. Spark for large-scale OO computation. Times in MM:SS.}
\label{fig:TPC}
\end{center}
\end{table}

Results are shown in Table \ref{fig:TPC}.  The difference in speed between the PC implementation and the Spark implementation
is significant.
When Spark data
are stored in
a hot HDFS, the two computations are $6\times$ to $66\times$ faster in PC.  
This is an apples-to-apples comparison, because in both systems, the data are being fetched from system storage, where they
can be buffered in OS buffer cache or PC buffer pool respectively.

If the Spark data are already
fully deserialized and stored as an RDD in memory, then PC is still 
between $1.5\times$ and $26\times$ faster
for both computations. Since in PC there is no distinction between
serialized and deserialized data, there is no analogous case in PC. 


One of the most striking---and surprising---results was that Spark had about the same performance for both computations.  This is
a bit surprising because the top-$k$ computation seems, on the face of it, much easier 
than the ``parts per supplier per customer'' computation.  $k$ was between 1,024 and 10,240, so (in theory) that should be
a hard limit on the number of customer's whose data are moved off of each machine during a shuffle 
(since it is impossible for more than $k$
customers processed on any machine to be in the top $k$).  One explanation could be not that Spark is surprisingly slow on the
top-$k$, but that PC is relatively slow on the ``parts per supplier per customer'' computation.  Profiling reveals that PC spends
a lot of time on \texttt{String} operations (looking up particular customers in the
\texttt{Map <String, Handle <Vector <int}\texttt{>}\texttt{>}\texttt{>}, for example).  
Because PC \texttt{String}s have the same representation in-RAM and on-disk, they are purposely designed to take little
space---they do not cache hash values, for example (unlike Java \texttt{Strings}).  This might explain why PC's speedup is
less significant on that computation.

\subsection {Machine Learning}

Finally, we consider three common machine learning algorithms: LDA,
GMM and $k$-means. 

\subsubsection {Implementations}

\noindent
\textbf{Latent Dirichlet Allocation.}
The first algorithm we implemented was a Gibbs sampler for
Latent Dirichlet Allocation, or LDA.  LDA is a common text mining algorithm.
While LDA implementations are common, we
chose a particularly challenging form of LDA learning:
a 
word-based,
non-collapsed Gibbs sampler \cite{jermaineExperimental}.  The LDA implementation
is \emph{non-collapsed} because it does not integrate out the word-probability-per-topic
and topic-probability-per-document random variables. In general, collapsed implementations
that do integrate out these values are more common, but such collapsed implementations
cannot be made ergodic in a distributed setting
(where ergodicity implies theoretical correctness in some sense).
Our implementation is
\emph{word based} because the fundamental data objects it operates over are \texttt{(docID, wordID, count)} 
triples.  This generally results in a more challenging
implementation from the platform's point-of-view because it requires a many-to-one join between words
and the topic-probability-per-document vectors.  In our experiments, there are approximately
700 million such triples, and each vector is around 1KB.  Hence, the many-to-one join between them results in
700GB of data.  If the platform does not manage this carefully, performance will suffer.

\begin{figure}
\centering
\includegraphics[width=0.5\textwidth]{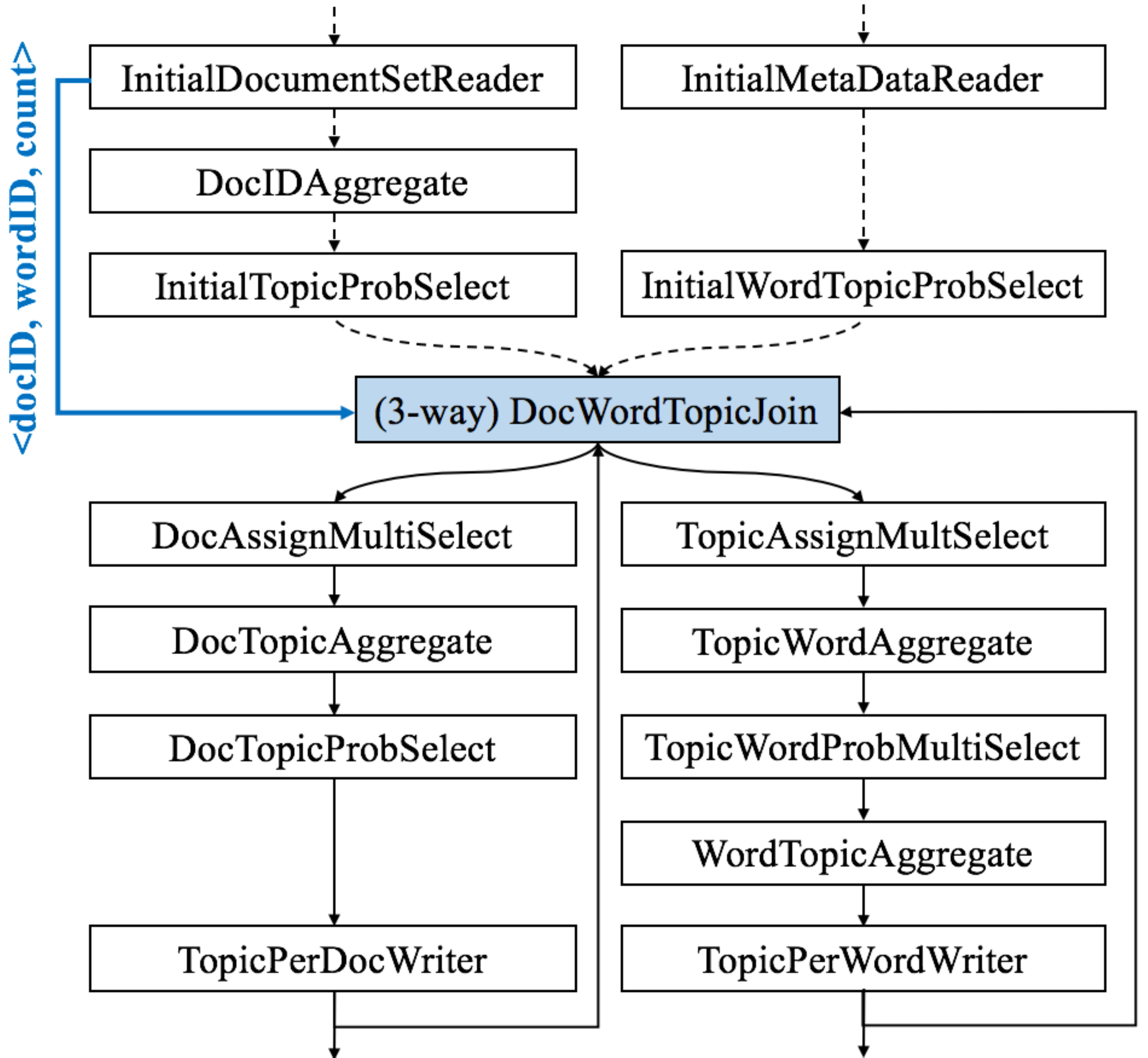}
  \caption{\label{fig:lda-query-graph} PC LDA's \texttt{Computation} objects and input-output dependencies. Computations
    connected by dash lines will only run once, during  
    initialization. Computations connected by solid lines will run iteratively.}
\end{figure}

The full PC LDA implementation requires fifteen different \texttt{Computation} objects, as shown in 
in Figure~\ref{fig:lda-query-graph}.  Each iteration requires a 
three-way \texttt{JoinComp}, three \texttt{MultiSelectionComp}s, and three
\texttt{AggregateComp}s, among others. 

Our PC LDA computation makes use of the GSL library~\cite{gsl} to perform all necessary random sampling (non-collapsed LDA requires 
sampling from both Multinomial and Dirichlet distributions).

We wished to compared our LDA implementation with an algorithmically equivalent Spark implementation.  Unfortunately,
Spark \texttt{mllib}'s LDA implementation is based on expectation
maximization (EM) and online variational Bayes.
Therefore, we had a Spark expert 
carefully implement an algorithmically
equivalent word-based, non-collapsed LDA Gibbs sampler on top of Spark.  His implementation used both
Spark's RDD and Dataset        
APIs as appropriate.
The required statistical computations use the
\texttt{breeze} package.

\vspace{5pt}
\noindent
\textbf{Gaussian Mixture Model.} A Gaussian mixture model (GMM) is a generative, statistical model where a dataset is modeled
as having been produced by a set of Gaussians (multi-dimensional Normal distributions). Learning a GMM using
the expectation maximization (EM) algorithm is one of the classical ML algorithms.
EM is particularly interesting for a distributed benchmark because in theory, the running time should be dominated
by linear algebra operations (such as repeated vector-matrix multiplications).

All linear algebra in our EM-on-PC implementation was performed using
the GSL library~\cite{gsl}.  
Our PC implementation uses a single \texttt{AggregateComp} object, which contains inside of it the current
version of the learned GMM model.  As this  \texttt{AggregateComp} is executed, a soft assignment of each
data point to each Gaussian is performed, and based off of this assignment, updates to each of the Gaussians
are accumulated.  The result of the aggregation is sent back to the \texttt{main} program where the actual update
to the model happens; the result is broadcasted in a new \texttt{AggregateComp} object, and the process begins again.

It turns out that an algorithmically equivalent implementation exists in Spark \texttt{mllib}.
Both implementations even use the same random initialization algorithm.
There are only slight differences between the two; for example, PC computation uses the standard ``log space'' trick to
compute the soft assignment and avoid underflow, whereas \texttt{mllib} uses thresholding.  

\vspace{5pt}
\noindent
\textbf{$k$-Means.} Our final ML benchmark algorithm is $k$-means.  
While not a particularly interesting computation, it is a now-classic
benchmark for Big Data ML.  We specifically developed our PC $k$-means implementation to closely match
the implementation in Spark's \texttt{mllib}.
Both implementations use the standard trick, where, to find the centroid closest to a given point,
a lower bound $||a - b||_2 \geq  abs(||a||_2 - ||b||_2)$ is
first computed to avoid unnecessary distance computations. 

\subsubsection {Experiments}

On the aforementioned eleven-node cluster, 
we ran all ML experiments using PC and Spark 2.1.0.

For LDA,  we
created a semi-synthetic document database with 2.5 million documents from
20-Newsgroups dataset by concatenating random pairs of newsgroup postings
end-on-end. There are more than 739
million \texttt{(docID, wordID, count)} triples in the dataset.
We use a dictionary size
of 20,000 words and a model size of 100 topics. 

For GMM, we generated
random data for three test cases: $10^7$ data
points with 100 dimensions, and $10^6$ data points with 300 and 500
dimensions, respectively. For each test case, the same random data was used
for comparing PC and Spark performance. For $k$-means, we
generate random data for $10^9$ data
points with ten dimensions, $10^8$ data points with 100 dimensions,  and $10^7$
data points with 1000
dimensions. Again, the same data is used on both PC and
Spark platforms.
Ten Gaussians are used for GMM, and ten clusters for $k$-means.

For each experiment, we carefully tune Spark
partition size, executor heap size and the number of cores to obtain maximum performance.
Input data for Spark are serialized
using Kryo, and read through a binary format (Parquet for the Dataset API,
and an Object file for the RDD API).

\subsubsection {Results and Discussion}

LDA Results (per iteration) are illustrated in Table~\ref{fig:LDA}.

\begin{table}[h!]
\begin{center}
\begin{tabular}{|c||c|c|c|c|c|c|}
\hline
PlinyCompute & \makecell{Spark 1: \\vanilla} & \makecell{Spark 2: also with \\join hint} & \makecell{Spark 3: also with \\forced persist} & \makecell{Spark 4: also hand-\\coded multinomial} \\
\hline
02:05 & 50:20 & 17:30 & 09:26 & 05:26 \\
\hline
\end{tabular}
\caption{PlinyCompute vs. Spark for LDA. Times in MM:SS, averaged over five iterations.}
\label{fig:LDA}
\end{center}
\end{table}

While Spark performed well, the 
amount of work required to arrive at a good solution 
was significant, representing about a week of tuning.  First, among other things, our Spark expert had to force a 
broadcast join.  Then, it was necessary to force Spark to
persist the result of one of the joins for later use.  Finally, it was necessary to hand-code a 
Multinomial sampler (avoiding the use of \texttt{breeze} for
multinomial sampling) to obtain an implementation that was competitive with PC.
This last bit of tuning (of course) can't be blamed on Spark, but the experience overall is illustrative: forcing 
a particular join and forcing a particular persist are \emph{workload specific} optimizations.  They may work for one
workload but be a poor choice for another, and require a tool end-user to actually change library code to achieve
performance.  In contrast, like a database system, PC is fully declarative in-the-large.
Decisions such as using a broadcast join instead of a full hash
join as well as which intermediate results to materialize (and which to pipeline or discard) are fully automated.

The results for GMM are illustrated in Table~\ref{fig:Gmm}. Here, PC achieved a 
$3\times$ speedup compared with Spark \texttt{mllib}'s GMM implementation
(using RDD APIs) for all cases.  We will discuss the significance of this finding in the next subsection, where we discuss
some of the issues surrounding Java vs. C++.  

\begin{table}[h!]
\begin{center}
\begin{tabular}{|c||c|c|c|}
\hline
Dimensionality & $100$ & $300$ & $500$ \\
Number of points & $10^7$ & $10^6$ & $10^6$ \\
\hline
\hline
PlinyCompute &00:30 & 00:38 & 1:42 \\
Spark \texttt{mllib} &1:41  &1:54 &5:05 \\
\hline
\end{tabular}
\caption{PlinyCompute vs. Spark for GMM. Times in MM:SS, averaged over five iterations.}
\label{fig:Gmm}
\end{center}
\end{table}

As illustrated in Table~\ref{fig:KMeans}, for $k$-means, PC achieved a $2\times$ to
$4\times$ speedup compared with the Spark \texttt{mllib} RDD implementation.
Curiously, the Spark \texttt{mllib} Dataset implementation
had performance similar to the RDD implementation for
$10^7$ data points and $10^8$ data
points, but much slower for $10^9$ data points. It turns out that 
the Spark \texttt{mllib} Dataset implementation first reads the data 
from a parquet file in the libSVM format, storing the data in a Dataset.  But then, the data
are converted into an RDD for processing, likely due to the relatively inflexible Dataset API.
This conversion becomes a bottleneck for the largest datasets.

\begin{table}[h!]
\begin{center}
\begin{tabular}{|c||c|c|c||c|c|c|}
\hline
& \multicolumn{3}{c||}{Initialization Latency} & \multicolumn{3}{c|}{Average
                                         Iteration Latency} \\
\hline
Dimensionality & $10$ & $100$ & $1000$ & $10$ & $100$ & $1000$\\
Number of points & $10^9$ & $10^8$ & $10^7$ & $10^9$ & $10^8$ & $10^7$\\
\hline
PlinyCompute &3:59 & 1:12 & 00:57 &00:37 & 00:09 & 00:06\\
Spark \texttt{mllib} RDD API &9:06  &4:18 &3:20 &01:02 & 00:28 & 00:23\\
Spark \texttt{mllib} Dataset API &15:12  &4:00 &3:07 &01:43 & 00:25 & 00:22\\
\hline
\end{tabular}
\caption{PlinyCompute vs. Spark for $k$-means. Times in MM:SS, averaged over five iterations.}
\label{fig:KMeans}
\end{center}
\end{table}

\begin{table}[h!]
\begin{center}
\begin{tabular}{|c||c|c|}
\hline
Applications & SLOC on PlinyCompute & SLOC on Spark\\
\hline
\texttt{lilLinAlg} &3505& 3130 (Scala)\\
TPC-H \texttt{Customer}s per \texttt{Supplier}&929 &953 (Java)\\
TPC-H top-$k$ Jaccard &793 & 966 (Java)\\
LDA &1038  &343 (Scala with \texttt{breeze})\\
GMM&932 & 474 (Scala with \texttt{breeze})\\
$k$-means &695  &670 (Scala)\\
\hline
\end{tabular}
\caption{PlinyCompute vs. Spark: lines of source code comparison.}
\label{fig:LOC}
\end{center}
\end{table}

\begin{table}[t]
\begin{center}
\begin{tabular}{|c||c|c|c|}
\hline
Matrix Dimensions & $1000\times1000$ & $10000\times10000$ \\
\hline
\hline
GSL &1033 ms &26:18  \\
Eigen &123 ms  &3:57\\
\texttt{breeze}-native &179 ms  &3:40\\
\hline
\end{tabular}
\caption{Single thread matrix multiplication benchmark tested in
  \texttt{m2.4xlarge} instance by setting thread number  to one for all
  packages. It shows that
  Java is as fast as C++ through invoking native code. (for
  $1000\times1000$ matrix multiplication, processing times are recorded in
  milliseconds, and for $10000\times10000$ matrix multiplication, processing times are
  recorded in hh:mm:ss format.)}
\label{fig:matrixMult}
\end{center}
\end{table}

\subsubsection{Experiments: Final Thoughts}

The central hypothesis in this paper was that ``declarative in the large, high-performance in the small'' can result
in an excellent platform for tool and library development.  We believe that these experiments have showed that.  The
most convincing benchmark was likely the first one, where 1.5 person-months of engineering time resulted in a tool
(the \texttt{lilLinAlg} tool for distributed linear algebra) that 
was faster than other competing systems with many years of development time behind then.  One of those systems (SciDB)
was implemented natively in C++, while the others (Spark's \texttt{mllib} as well as \texttt{SystemML}) used Hadoop and Spark.  
Other benchmarks showed similar advantages to PC, with complex object manipulations being 
up to $66 \times$ faster on PC than on Spark, and ML computations generally being around $3 \times$ faster on PC than on Spark.

We close the benchmarks with two final questions.  First: PC may be faster, but is it significantly more difficult to develop
for than a platform that uses a managed runtime?  PC certainly gives a programmer more flexibility, and with that can come
certain costs---less knowledgeable developers may find PC difficult to code for.  But, by at least one metric--source
lines of code (SLOC)---PC is \emph{not}
any more difficult as a development target than Spark.
Table~\ref{fig:LOC} shows the SLOC counts for the various implementations described here, comparing to their Spark counterparts.
If one believes that engineering effort is roughly proportional to SLOC written, there is not a significant
difference between the two
systems.  
While for LDA and GMM PC required $2\times$ to $3\times$ the code required for Spark, a lot of that was related
to the fact that Scala has a nicer interface to numerical routines (via \texttt{breeze}) 
than does GSL, which was used in those
implementations.

Our last question is: PC may be faster, but how much of that is related to ``declarative in the large, high performance in 
the small?"   Isn't C++ simply faster than Java, and might that explain a lot of the advantage realized by PC?  We begin
our answer to this question by pointing out that SciDB is written in C++, and not Java, so the C++ vs. Java question is not 
relevant to all of our findings.
But even when the question is relevant, we assert that the most significant
difference between Java on a modern JVM and C++ is that the latter
gives the system developer more control over issues such as memory management, 
which the developer may use to produce a faster system (this is precisely
what we have attempted to do with our development of the PC object model, for example).  
There is nothing inherent in C++ that makes it faster than Java if this extra flexibility is not used properly, 
especially in the age of
JIT compilation and generational garbage collectors.  

In fact, there is some good evidence that Spark and Java may have had some significant
built-in \emph{advantages} vs. our C++ implementations.
Out of curiosity, we ran a simple micro-benchmark on an AWS \texttt{m2.4xlarge} machine, where we compared the various
packages for statistical/scientific computing used throughout these experiments.
In this benchmark, we run a single-thread matrix multiplication
to compare Java \texttt{breeze} (used in all Spark implementations) 
with Eigen (used by PC's \texttt{lilLinAlg}) and GSL (used in all of our PC ML implementations).
The results are shown in Table~\ref{fig:matrixMult}.  
Here we find that Java Breeze has slightly better performance than Eigen and \emph{much}
better performance than GSL.  
Thus, in one way, our C++ implementations were at a significant disadvantage compared to Java.

The point is that achieving excellent performance on complex, distributed computations is never a simple matter
of ``use C++, not Java''~\cite{ousterhout2015making, shi2015clash}.  Many factors go into having a superior implementation, and those tend to even things out,
as some of those factors 
go \emph{against} PC.  We argue that the reason that PC was consistently faster
than its competitors are the design principles underlying the system---which indeed was enabled by the choice of language---and 
not the programming language itself.

%% file: rel_work.tex
\section{Related Work}
\label{sec:survey}
The PC project was inspired by works from database engines, distributed in-memory processing
frameworks,  programming languages and compilers, which can be
categorized into two classes for convenience of discussions: systems and techniques with managed
run-times, and without managed run-times.

\subsection {Systems and Optimizations for Managed Run-times}

Most recent big data query processing and analytics
systems are implemented using
high-level programming languages such as Java and Scala  ~\cite{dean2008mapreduce, yu2008dryadlinq,
  neumann2011efficiently, zaharia2012resilient,
  alexandrov2014stratosphere, klonatos2014building,
  crotty2015tupleware, armbrust2015spark} which rely on a managed
runtime such as the JVM for object (de-)allocation and memory and virtual function call management. Numerous papers
have looked at the problem of mitigating the costs incurred, and propose techniques aiming at
reducing the overhead incurred by the managed runtime.  Ideas examined include
code generation, garbage collector (GC) tuning, and the
use of off-heap memory and structured objects.
We briefly discuss some of those efforts now.

\vspace{5pt}
\noindent
\textbf{Code Generation.} The basic idea is that a system implemented in high level language
can generate native query execution code that manually manages
memory and also avoids virtual function call
overheads. DryadLinq~\cite{yu2008dryadlinq} allows a user to express
distributed data flow
computations in a high-level language like C\# and strongly typed .NET
objects, and it compiles those computations into .NET assembler.
LegoBase~\cite{klonatos2014building} switches the interface
from declarative SQL to a high-level language (Scala) and uses a query engine
written in Scala as a code generator to emit specialized and low-level
C code for execution. TupleWare~\cite{crotty2015tupleware} supports
multiple high-level languages (any language with an LLVM compiler) 
and aims to
optimize for UDFs by utilizing code
generation to integrate UDF code with the engine 
code. 
Weld~\cite{palkar2017weld} is a recent system developed in Scala and
Python. It proposes
a common runtime for data analytics libraries by asking library
developers to express their work using a new intermediate
representation (IR) and compiles this IR into multi-threaded code using
LLVM.  Then, application developers can use unified APIs to
call different libraries from Weld. Since version 2.0, Spark~\cite{zaharia2012resilient}
also exploits whole-stage code generation to generate JVM 
code.  The goal is mainly to reduce type parsing and virtual function call
overhead. PC uses a form of code generation (template metaprogramming) but 
the emphasis is quite different, in the sense that the goal is to allow for
efficient distributed programming with complex objects.

\vspace{5pt}
\noindent
\textbf{Optimized Memory Management.} 
Apache Flink~\cite{alexandrov2014stratosphere} and Apache
Spark~\cite{zaharia2012resilient} are both distributed in-memory
dataflow systems. They both provide high level language interfaces (Java and Scala).
Spark SQL~\cite{armbrust2015spark} is a relational system built on top
of Spark. These systems all attempt to 
alleviate garbage collection overhead by storing data
in untyped byte arrays or even storing data off-heap.

Apache Flink~\cite{alexandrov2014stratosphere} in particular
assigns memory budgets to its data processing operators. Upon
initialization, a computation requests a memory budget from the
memory manager and receives a corresponding set of byte arrays as memory segments. 
Each operation will then have its own memory pool that it can manually
manage. 
Apache SparkSQL~\cite{armbrust2015spark} serializes 
relational tables into byte arrays and stores the serialized bytes
in a main-memory columnar storage. Spark Tungsten~\cite{tungsten}
optimizes the Spark execution backend by grouping execution
data (such as hashed aggregation data) 
into byte arrays and data can be allocated off-heap via
the sun.misc.Unsafe API, reducing
GC overhead. Deca~\cite{lu2016lifetime} is a memory management framework aiming at
reducing GC overhead. It stores
various Spark data types, e.g. UDF variables, user data and
shuffle data into different
off-heap containers so that objects in each container can have a similar
lifetime and can be recycled together.
All of these methods attempt to alleviate GC overhead; in contrast, PC simply does
not use a managed runtime.

\vspace{5pt}
\noindent
\textbf{Relational Processing on Binary or Structured Objects.} The idea here is to convert or
serialize a Java Object into an efficient binary representation and directly
operate on this binary representation, thus reducing serialization and
deserialization overhead.  This 
binary data representation may also be more space efficient. In addition, binary objects
with similar lifetimes can be easily stored in consecutive byte
arrays, significantly reducing GC overhead.

Apache Flink~\cite{alexandrov2014stratosphere} uses reflection
to analyze Java/Scala object types, and it
maps each object type to one of a limited set of
fundamental data types, such as Java primitive types, an array type,
a Hadoop Writable type, a Flink fixed-length tuple, and so on. Each
fundamental data type is associated with a serializer.  Certain data
types provide comparators to efficiently compare binary
representations and extract fixed-length binary key prefixes without
deserializing the whole object.

Spark~\cite{tungsten} has introduced the Dataset/Dataframe representations
to complement the more object-oriented
RDD representation~\cite{zaharia2012resilient}. Datasets/Dataframes are
binary data representations used to encode JVM objects relationally (as a
row with various fields/columns). 
Datasets/Dataframes enable relational-style processing
through a relational query optimizer called Catalyst and
also enables Java intermediate code generation to reduce virtual
function call overhead through Tungsten~\cite{tungsten}. 

Such techniques significantly boost performance, by moving away from a flexible, object-oriented
type of system to a more relational system.
It is known that relational systems can be fast, but they limit the sort of applications that
can easily be coded on top of the system.  In contrast, PC attempts to offer a fully 
object-oriented interface.

\subsection {Related Native Systems}
Despite the decreasing performance gap between Java/Scala and C/C++,
operating systems and many tools or libraries are still developed in C/C++. For example,
all of the popular numerical processing libraries utilized in our benchmark are
implemented in C/C++. 
While it is rarer for modern Big Data systems to be implemented natively, some systems do exist.
We now categorize and describe such systems.

\vspace{5pt} 
\noindent
\textbf{HPC Systems.} HPC systems such as Charm++~\cite{kale1993charm++}, OpenMP~\cite{dagum1998openmp}, Cilk~\cite{blumofe1996cilk},  Intel's Array Building
Blocks~\cite{newburn2011intel}, and Threading Building Blocks~\cite{reinders2007intel} are built on
low-level interfaces such as MPI~\cite{gropp1996high}.
Systems built using these tools can be very fast.  However, as described in the introduction to the paper, they are not often used
for modern Big Data programming.

\vspace{5pt} 
\noindent
\textbf{Query Processing.} Many of the ideas underlying PC's vectorized processing engine
were pioneered in various relational systems.  Vertica/C-Store~\cite{stonebraker2005c} proposes
to store and process relational tables by
column to optimize for read performance.
VectorWise~\cite{zukowski2012vectorwise}, which grew out of the MonetDB/X100 project, makes use of
vectorized query execution to amortize the virtual function call overhead across records---just like PC---and also
to exploit SIMD support
on modern hardware. Voodoo~\cite{pirk2016voodoo} uses a new
declarative algebra as the compilation target for query plans, to
enable tuning for hardware specifics like caches, SIMD registers,
GPU and so on. It requires a relational front-end, and serves as an alternative backend for
relational databases such as MonetDB and Hyper. While PC TCAP compilation
and processing is closely related to prior ideas from work in 
code generation and vectorized processing (and PC could, for example,
leverage Voodoo's hardware tuning ideas), PC is non-relational, and attempts to
facilitate high-performance processing on top of a 
highly expressive object model.

\vspace{5pt} 
\noindent
\textbf{Cloud Frameworks.} Impala~\cite{bittorf2015impala} is a
C++-based 
SQL query engine that relies on Hadoop for scalability and 
flexibility in interface and schema. Impala compiles SQL 
into LLVM assembler.
However,
Impala uses a relational data model (though it can read/write
semi-structured data in storage formats such as Arvo, Parquet, RC and so
on from/to external storage such HDFS, using standard
serialization/deserialization methods).

Spanner~\cite{bacon2017spanner} is a
distributed data management system that backs various operational services
at Google. Spanner started out as a key-value store and evolved into a
relational database system with a SQL query processor. It implements a
dialect of SQL, called \emph{standard SQL}, which uses arrays and structures to
support nested data as a first class citizen. To integrate
with user applications, the relational data described by standard SQL needs to
be translated to protocol buffers or user languages through the
GoogleSQL library that contains a compiler front-end and a library of
scalar functions. PC takes a fundamentally different approach, as all code is 
object-oriented rather an a mix of SQL and other high (or medium) level languages.

Tensorflow~\cite{abadi2016tensorflow} is a
distributed computing framework mainly designed for deep learning. It
mainly supports processing of numerical data with a
very limited set of types.
Tensorflow provides
a much lower level API than PC's declarative interface, based on tensors,
variables and sessions.

%% file: conc.tex
\section{Conclusions}
\label{sec:conc}
This paper has described PlinyCompute, or PC for short.  PC is a system for the development of high performance
distributed data processing tools and libraries.  PC is designed to inhabit the space between
high-performance computing platforms such as OpenMP and MPI, which provide little direct support
for managing very large data sets, and dataflow platforms such as Spark and Flink, which rely on 
a managed runtime to provide low-level services such as allocation and deallocation of data objects and memory management.
PC's guiding design principle is ``declarative in the large,
high performance in the small''.
In the large, PC presents the capable systems programmer with a very high-level, declarative interface, relying on automatic,
relational-database style optimization.  This is crucial for tool and library development, since the same tool should run well
regardless of the characteristics of the data and of the compute platform.  But in the small, PC relies on the PC object model.
The PC object model is an
API for storing and manipulating persistent data, and has been co-designed with PC's memory management
system and computational engine to provide maximum performance.
One of the key ideas behind the PC object model is the \emph{page as a heap principle}. All PC
Objects are allocated and manipulated in-place, on a system- (or user-) allocated page. There is no distinction
between the in-memory representation of data and the on-disk (or in-network) representation of data.
Thus there is no (de-)serialization cost to move data to/from disk and network, and memory management
costs are very low.

We have performed a reasonably extensive set of benchmark experiments that indicate that these ideas can result in 
a system that is very high performance and yet offers a relatively simple and usable object oriented API.  In particular, 
we have given strong evidence that PC \emph{is} particularly well-suited to tool and library development. 
For example, we asked a PhD
student (who at the outset knew nothing of PC) to use the system to build a small Matlab-like programming
language and library for distributed matrix operations called
\texttt{lilLinAlg}.  The resulting tool outperformed many other, long-lived tools
for the same purpose, such as SciDB, \texttt{mllib}, and SystemML.  SystemML in particular has resulted 
in several research papers,
including one awarded a VLDB best paper award \cite{boehm2016systemml}.
\texttt{lilLinAlg} was developed in six weeks by a single PhD student.
One may conjecture that had SystemML been built on a platform such as PC rather than on Spark, it might be significantly
faster than it is now.

Many avenues for future work remain such as cost-based query
optimization and more specialized support for graph processing, deep
learning and so on. Also there is significant scope for expanding the functionality and
usability of the PC object model, including relaxing the strict requirement that all \texttt{Handle}s point within a page.  This
would allow pointer-like objects to point off a page, perhaps even off a machine.

%% file: example.tex
\section{$k$-means Example}
\label{sec:example}
Imagine that a user wished to use PC to build a high-performance
library implementation of a $k$-means algorithm \footnote{The $k$-means implementation
    described in this section is
    different with (and much simpler than) the implementation we used
    for benchmark in Section~\ref{sec:exp}.}.
Once the programmer had defined the basic type over which the clustering is to be performed (such as the
\texttt{DataPoint} class), 
a programmer would likely next define a simple class that allows the averaging of vectors:

\begin{codesmall}
class Avg : public Object {
	long cnt = 1;
	Handle <Vector <double>> data = nullptr;
	Avg &operator + (Avg &addMe)
           {/* add addMe into this */}
};
\end{codesmall}

\noindent
The programmer might next add a method to the \texttt{DataPoint} class that converts the \texttt{DataPoint} object to an \texttt{Avg} object:

\begin{codesmall}
Avg DataPoint :: fromMe () {
	Avg returnVal;
	returnVal.data = data;
	return returnVal;
}
\end{codesmall}

\noindent
And also add a method to the \texttt{DataPoint} class that accepts a set of centroids, computes the Euclidean distance to
each, and returns the closest:

\begin{codesmall}
long DataPoint :: getClose (Vector <Vector <double>> 
        &centroids) {...}
\end{codesmall}

\noindent
Next, a programmer using PC would define an \texttt{AggregateComp} class using PC's lambda calculus, since, after all, the $k$-means algorithm is essentially 
an aggregation:

\begin{codesmall}
class GetNewCentroids : public AggregateComp 
    <Centroid, long, Avg, DataPoint> {

public:
   Vector <Vector <double>> centroids;

   Lambda <long> getKeyProjection (
       Handle <DataPoint> aggMe) override {
          return makeLambda (aggMe, 
              [&] (Handle <DataPoint> &aggMe) 
                 {return aggMe->
                     getClose (centroids);});
   }
   Lambda <Avg> getValueProjection (
       Handle <DataPoint> aggMe) override {
          return makeLambdaFromMethod 
              (aggMe, fromMe);
   }
};
\end{codesmall}

\noindent 
The declaration \texttt{AggregateComp <Centroid, long, Avg, DataPoint>} means that this computation aggregates
\texttt{DataPoint} objects.  For each data point, it will extract a key of type \texttt{long}, a value of type \texttt{Avg}, which will be
aggregated into objects of type \texttt{Centroid}.  To process each data point, the aggregation will use the lambdas constructed by
\texttt{getKeyProjection} and \texttt{get ValueProjection}.  
In this case, for example,
\texttt{getKeyProjec tion} builds a lambda, which simply invokes the native C++ lambda given in the code---this
native C++ lambda returns the identity of the centroid closest to the data point.

To build a computation using this aggregation class, a programmer would need to specify the \texttt{Centroid} class (the result of this aggregation):

\begin{codesmall}
class Centroid : public Object {
	long centroidId; 
	Avg data;
public:
	long &getKey () {return centroidId;}
	Avg &getVal () {return data;}
};
\end{codesmall}

\noindent
And then build up a computation using these pieces:

\begin{codesmall}
Handle <Computation> myReader = 
    makeObject <ObjectReader <DataPoint>>
     ("myDB", "mySet");
Handle <Computation> myAgg = makeObject 
    <GetNewCentroids> ();
myAgg->centroids = ... // initialize the model
myAgg->setInput (myReader);
Handle <Computation> myWriter =  makeObject <Writer
     <Centroid>> ("myDB", "myOutSet");
    myWriter->setInput (myAgg);
pcClient.executeComputations (myWriter);
\end{codesmall}

\noindent After execution, the set of updated centroids would be stored in \texttt{myDB.mySet}.
Performing this computation in a loop, where the centroids are repeatedly updated until convergence, completes the implementation.

%% file: allocation.tex
\section{Object Model Tuning}
\label{sec:allocation}

\noindent
The PC object model is designed for zero-cost data movement, the result being that there is often no serialization or deserialization
cost
when moving PC \texttt{Object}'s across processes.  But memory management can still be costly.  Deallocating and cleaning
up complex objects (in particular, instances of container classes) can require significant CPU resources, which, depending upon the 
circumstance, may be un-necessary.  In-keeping with the assertion that application programmers should be in
control of performance-critical policies, it is possible to explicitly control how memory is reclaimed and re-used during PC computations.
This is facilitated through a set of \emph{allocation policies} that a programmer can choose from.

When the reference count for a PC \texttt{Object} located in a managed allocation block goes to zero, it is deallocated.  The exact
meaning of ``deallocated'' is controllable by the programmer, via a call to the \texttt{setAllocatorPolicy} on each computation object
that is created (\texttt{JoinComp}, \texttt{SelectionComp}, etc.).  Currently, PC ships with
three allocator policies:

\begin{enumerate}

\item Lightweight re-use.  This is the default policy.  When a PC \texttt{Object} is deallocated, its space in the allocation block is made available for re-use by
adding the space to a pool of similarly-sized, recycled memory chunks (all recycled chunks are organized into buckets, where a chunk of size
$n$ goes into bucket $\log_2 (n)$).  A request for RAM in a block is fulfilled by first scanning the recycled chunks in the appropriate bucket, then
attempting to allocate new space on the end of the block, if that fails.
\item No re-use.  The space containing deallocated PC \texttt{Object}s is not-reused.  Hence, it is very similar to classical, region-based allocation---though PC \texttt{Object}s
are reference-counted, and a destructor is called for each unreachable PC \texttt{Object}.
On the positive side, this allocation policy is very fast.  On the negative side, frequent allocations of temporary PC \texttt{Object}s will result in a lot of wasted space.
\item Recycling.  This is layered on top of lightweight re-use.  When the recycling allocator is used, any time a fixed-length
PC \texttt{Object} is deallocated, it is
added to a list of objects all having the same type.  All calls to \texttt{makeObject} with the zero-argument constructor will
pull an object off of the list of recyclable objects for the appropriate type.  
If an object is available for recycling, it is returned.  If not, or if any other constructor other than the zero-argument constructor is called, 
then the lightweight re-use allocator is used to allocate space for the requested
object.

\end{enumerate}

Note that
variable-length objects are never recycled.  There are just a few of these types in PC, and they are typically
used internally to implement the built-in PC container
types, and not by PC application programmers.  For example, PC's variable-length
\texttt{Array} class is used to implement the standard PC \texttt{Vector} container.  
These are not recycled because recycling allocations of such objects would need to match both on type and on size.  Matching on both at once would 
be computationally expensive, and could also allow long lists of objects to build up, waiting to be re-used.

In addition to policies that can be set on a per-computation basis,
it is also possible for a programmer to supply the following policies, on a per-\texttt{Object} bases, during PC \texttt{Object} allocation:

\begin{enumerate}

\item No reference counting.  This PC \texttt{Object} is not reference counted, and it is not included in the total count of objects on an allocation
block.  If each PC \texttt{Object} on an allocation block
is allocated in this way, this results in pure, region-based memory management, and is exceedingly lightweight.
\item Full reference counting.  This is the default.
\item Unique ownership.  The PC \texttt{Object} is not reference counted, but there can be one \texttt{Handle} object referencing the uniquely-owned
object.  When that \texttt{Handle} is destroyed, the object is deallocated.

\end{enumerate}

%% file: pipelined_engine.tex
\begin{figure}[t]
  \begin{center}
    \includegraphics[width=6in]{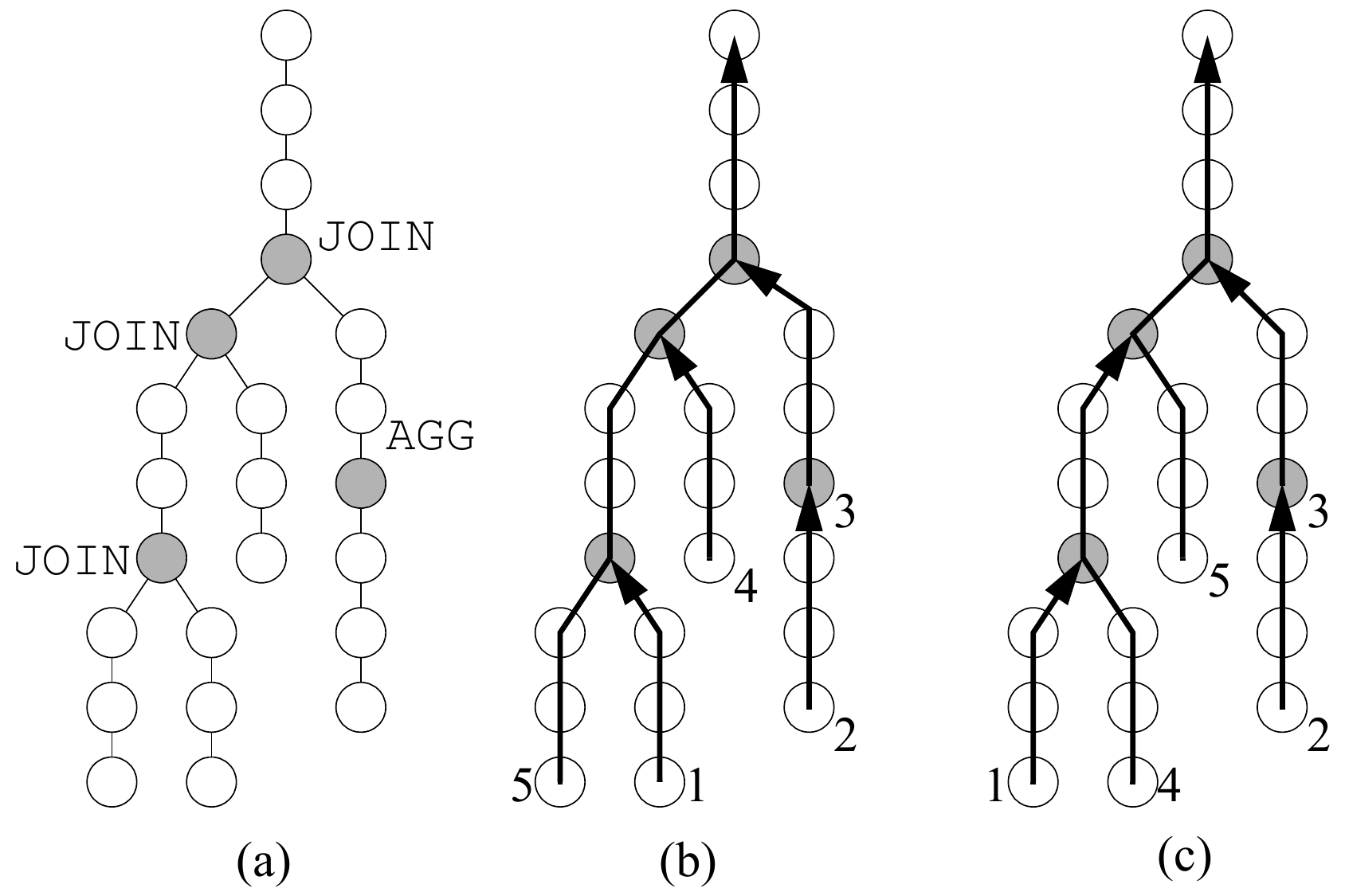}
  \end{center}
\vspace{-10 pt}
  \caption{TCAP code containing three joins and an aggregation represented as a graph (a).  During physical planning, it must be broken into pipelines; two
potential pipelinings are shown in (b) and (c).}
  \label{fig:TCAP}
\end{figure}

\section{Pipelined Execution Details}
\label{sec:pipelined}

This section describes some of the details regarding PC's pipelined execution engine.  We begin by considering how a TCAP
program is used to build a set of pipelines, and then focus on interaction between pipelined execution and the PC object model
and memory management. The description in this section is mainly focused on a
single-thread scenario. We will introduce parallel and distributed
processing in Section~\ref{sec:implementation} of the Appendix.

\vspace{5 pt}
\noindent
\textbf{Breaking a TCAP DAG into Individual Pipelines.}
The previous section discussed the problem of logical optimization for TCAP programs---optimization that takes into account the semantics
of the various operations, but does not consider actual implementation.  
Executing a TCAP program using PC's vectorized execution engine requires physical planning as well.  The single most important physical 
optimization involves choosing how to break a TCAP program into a set of individual \emph{pipelines}, and then choosing an execution order
for those pipelines.
A pipeline is a list of TCAP operations that are executed in sequence over a vector list, with each operation performing some
transformation over the vector list; at all times as the vector list is pushed through the pipeline, it remains buffered in RAM.  
At the end of a pipeline is a \emph{pipe sink}, where one or more of the
vectors in a vector list is written out to a PC \texttt{Object}
container located on an output page, for later use (which may require shuffling or broadcasting the container across a PC cluster).  
For example, the pipe sink associated with aggregation
writes a vector of key \texttt{Object}s and a vector of value \texttt{Object}s to a PC \texttt{Map}, performing pre-aggregation when
applicable.
Each pipeline ends in a pipe sink, though only a few TCAP operations require pipe sinks: \texttt{JOIN} (where either one or both inputs are pipe
sinks, depending upon the join algorithm chosen), \texttt{AGG}, and
\texttt{OUTPUT}. In our current implementation, if one operation's
output has multiple consumers (the output is the input of more than one operations), the output will also be
materialized to a pipe sink.
Two different decompositions of a TCAP program into individual pipelines are shown in Figure \ref{fig:TCAP}.

\vspace{5 pt}
\noindent
\textbf{Ensuring In-Place Data Allocation of Output Data.}
Our primary goal when designing PC's pipelined execution engine is reducing  
memory-related costs.  
Since we 
encourage programmers to manipulate (potentially complex) PC \texttt{Object}s
rather than flat data, care must be taken to ensure that memory-related costs do
not dominate execution times.

Of all memory-related issues,  
data placement is paramount.  Because PC is designed to operate over
complex objects, data movement can be 
expensive.  To maintain the principle of
zero-cost data movement, \texttt{Handle}s pointing to out-of-block data are not allowed in PC.
Thus, copying a PC \texttt{Handle} from one memory block to another requires a deep copy of the target of the \texttt{Handle} to the new block.  
And since a
\texttt{Handle} may be declared as pointing to a super-type of its
target type (for example, an \texttt{Emp} object may be pointed to by a \texttt{Handle <Object>}), the correct code for invoking the deep copy of the
target must be executed via a virtual 
function call, which can in turn set off a cascade of virtual function calls. It is this chain reaction that we seek to avoid at all costs.
Thus, the most important principle governing the 
design of PC's pipelined execution engine is that \emph{data should be constructed where it is ultimately needed}.

To facilitate this, we note that user-supplied code will typically accept input data, then somehow use that input data to
create an output object.  For example, consider 
the following lambda term construction function, specifying the creation of output data:

\begin{codesmall}
Lambda <Handle <Emp>> getProjection (Handle <Sup> arg) {
        return makeLambda (arg, [] (Handle <Sup> &input) -> Handle <Emp> {
		return makeObject <Emp> (input->getName (), input->getDept ());
	});}
\end{codesmall}

\noindent
When the code corresponding to this lambda term is executed as a pipeline,
we wish to ensure that there is no need to move the \texttt{Emp} object created within the native C++ lambda to an output page.
To guarantee this,
PC obtains from the buffer pool a page for writing output objects for each thread, and places the current object allocation block
onto this page. 
When the call to \texttt{makeObject} within the lambda is executed, it will create the output \texttt{Emp} object
directly on the output page.  Any allocations triggered by the construction of the \texttt{Emp} will also be directly on the output page.

\vspace{5 pt}
\noindent
\textbf{Avoiding Unwanted In-Place Allocations.}
It is just as important to avoid unwanted allocations on an output page.  Depending upon the user-supplied
allocation policy, such allocations will either result
in holes on the page when it is shipped into the distributed PC cluster and hence a lower data density (if no-reuse is specified), 
or else they will result in the utilization of CPU cycles to 
reclaim and reuse the memory. 

Some of the burden of avoiding unwanted allocations is placed on a PC programmer. A programmer should
avoid unnecessary calls to \texttt{makeObject ()}, specifically avoiding those calls that allocate data that can never possibly make it to an output object.

Likewise, the system should do the same.  It is most critical that PC avoids moving intermediate data to an output page during pipeline execution.
Reconsider the example of Section \ref{sec:vectorized}.  The user-specified \texttt{getSelection ()} describes
a computation that invokes \texttt{Emp::getDeptName ()}, checking whether it equals the extracted value of
\texttt{Dep::deptName}.  During pipelined execution, care is taken \emph{not} to store these intermediate results
on the output page.  The vector that results from extracting
\texttt{Dep::deptName} from each input object is allocated outside of
the output page, and stores C-style pointers to each \texttt{Dep::dept
  Name}
value that are simply dereferenced when the vector is used by subsequent pipeline stages.  
If \texttt{getDept Name ()} returns a reference, its output is treated similarly.  If 
\texttt{getDeptName ()} returns actual data (and not a reference), then that data is stored in a vector \emph{outside}
of the output page.  Thus, in the case that the returned object has \texttt{Handle}s that refer back to the input data, a deep
copy will not be invoked because those \texttt{Handle}s will physically reside on a memory location outside of the current
allocation block (as described previously, deep copies of the target of a \texttt{Handle}
happen only when a \texttt{Handle} is copied to the current allocation block).  
It is only when/if a subsequent operation copies the result (and the \texttt{Handle}s are contained)
to the output page (corresponding to the current allocation block)
that a deep copy happens.  Thus, intermediate data are \emph{lazily copied} to 
the output page, once \texttt{Handle} to the data is copied to the output page, which tends to avoid un-necessary allocations.

\vspace{5 pt}
\noindent
\textbf{Memory Dependencies During Pipelined Processing.}
All data produced via calls to \texttt{makeObject} are written to
allocation blocks housed on in-memory pages served to PC by the buffer
pool.  PC's pipelined
processing induces four types of pages, each of which has a different lifetime during which it must be buffered in RAM:

\begin{enumerate}

\item The first page type is an \emph{input page} that stores the base data that will be processed by the pipeline.
To push vector lists into a pipeline, PC first loads a page containing a PC container (such as a PC \texttt{Vector}) of 
input data from the buffer pool, and then repeatedly creates vectors of \texttt{Handle}s to the objects on
the page (the little `v' is intentional; these vectors are \emph{not} PC \texttt{Object}s themselves, as we do not want them to be allocated
on the current  allocation block, as they store only intermediate data).  PC wraps
each vector produced in a vector list, and sends the vector list into the pipeline.  This process is repeated until the input page is consumed.
The input page must be buffered as long as a vector list originally created using its data is making its way through the pipeline.  

\item A \emph{live output page} that houses the current allocation block.  All allocations happen on this block, and so this page may hold intermediate data
and/or output data.
Intermediate data are 
reachable only by vectors that have been produced
by non-sink stages in the pipeline; as described above, the expectation is that intermediate data will eventually become output data stored on the
same page, avoiding a copy.  Output data are
data written by the sink (terminal) node in the pipeline to a PC container 
\texttt{Object} (such as a \texttt{Vector} or a \texttt{Map} if the pipeline is feeding a join) that holds the ultimate output
of the pipeline.

\item Zero or one \emph{zombie output pages} that are full and store output data, but must
be pinned in RAM because they also store intermediate data.  This may happen when one or more vector lists make it all the way through the pipeline,
causing output data to be written by 
the sink node at the end of the pipeline.  Then, the next vector list only makes it part way through the pipeline before the live
output page
fills and an out-of-memory fault occurs.  
At this point, a new live output page is obtained for use
as the current allocation block.  However, since the previous live output page still contains valid intermediate data, it cannot be written out.  This 
page becomes a \emph{zombie output page} and remains that way until it cannot possibly hold any intermediate data, at which point it can be flushed.

Note that there can be at most \emph{two} zombie output pages for a pipeline,
because a pipeline is restricted 
to have just one sink.  
When a zombie output page is created, the vector list whose processing is
responsible for its creation may create references to intermediate data allocated to the new, live output page, and then processing of that
vector list begins
to fill the live output page with output data.  This may induce a second
out-of-memory fault, creating a second zombie output page.  However, since the vector list whose processing induced the fault is being
processed by the pipeline's sink, \emph{it cannot be used to produce additional intermediate data whose references will be held by the vector list} 
and can only produce output data.  Thus, any additional pages the pipeline is used to produce
cannot contain a mix of output and intermediate data.  Further, note that once a vector list makes it all the way through the pipeline, all of its references
to data are destroyed, so no pages can possibly
store valid intermediate data and hence all zombie output pages can be flushed---and hence the number of zombie output pages is capped at two.

\item Finally, there are zero or more 
\emph{zombie pages} that store only intermediate data.  These are similar to zombie output pages, but they do not store any output data; they
store only intermediate data.  Just like zombie output pages, all zombie pages can be flushed whenever a vector list is completely processed by a pipeline.
However, unlike zombie output pages, zombie pages should not be written back.
\end{enumerate}

%% file: implementation.tex
\section{Implementation}
\label{sec:implementation}

Thus far, we have mostly considered lower-level systems issues (the design and implementation of the PC object mode, TCAP optimization and vectorized processing) and
the programming interface that PC offers to end-users
.  In this
section, we discuss how these pieces fit together into a PC
distributed cluster at large.  First we 
describe PC's overall architecture, and then we detail how all of these pieces work together to process distributed aggregations and distributed joins.


\subsection{System Architecture}

The overall PC system architecture is illustrated in Figure~\ref{fig:arch}.
A PC cluster consists of a \emph{master node} as well as one or more \emph{worker nodes}.
The master node runs a \emph{master server}, which consists of
four software components: 

\begin{figure}
\centering
\includegraphics[width=0.75\textwidth]{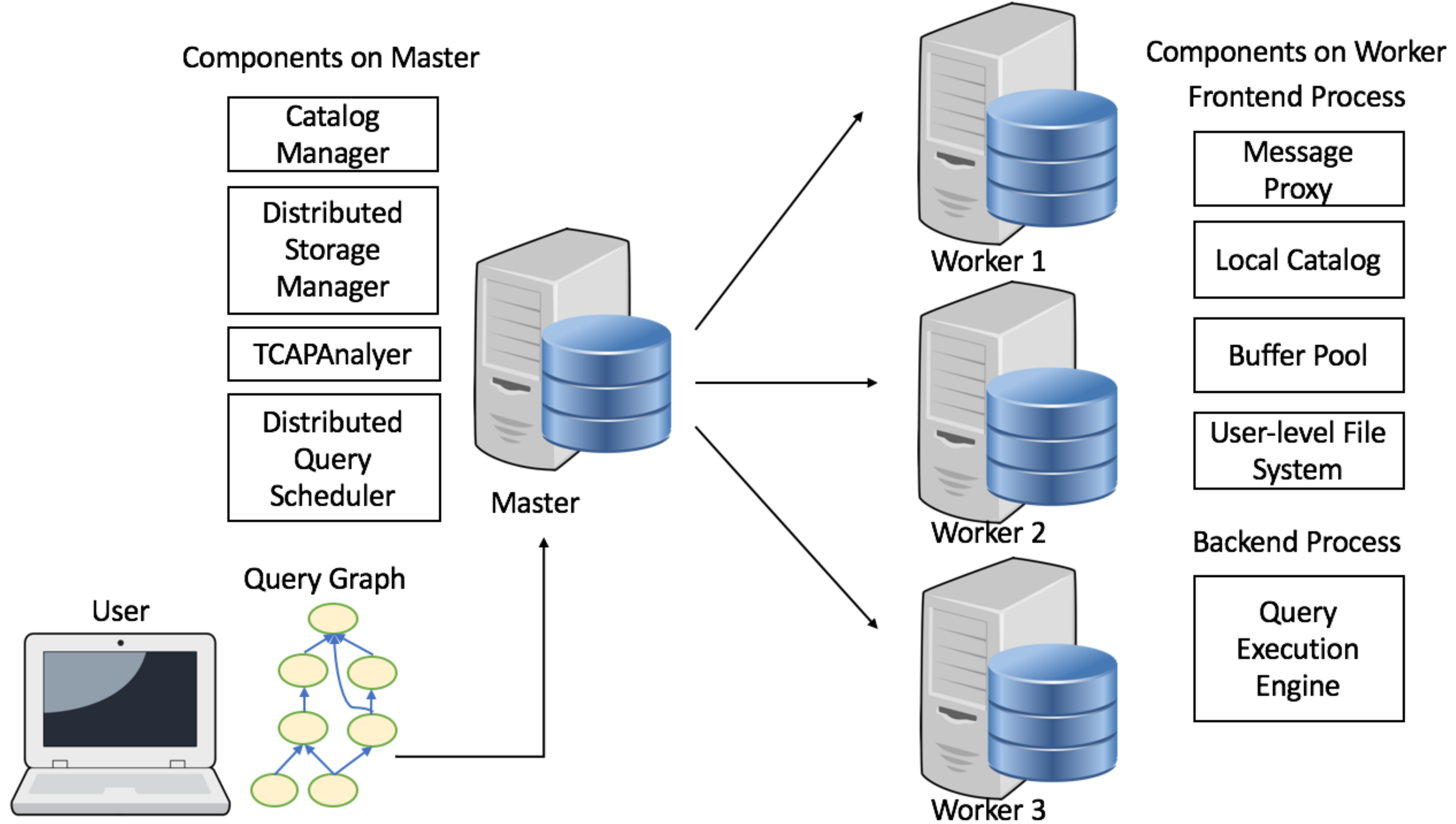}
  \caption{\label{fig:arch} PC distributed runtime.}
\end{figure}

\begin{enumerate}
\item The \emph{catalog manager}, serving system meta-data (including the master copy of
the mapping between type codes and PC \texttt{Object} types) and compiled, shared libraries
for performing computations over PC \texttt{Object}s;
\item The \emph{distributed storage manager}, 
the centralized server that manages PC's storage subsystem; 
\item The \emph{TCAP optimizer} which is responsible for optimizing
  programs that have been compiled into PC's domain-specific TCAP
  language (as in 
Section~\ref{sec:optimizer} of the paper);
\item The \emph{distributed query scheduler} that is responsible for accepting optimized TCAP computations.
It dynamically transforms those computations into a set of \texttt{JobStage}s,
such as:
\texttt{PipelineJobStage}, which consists of a series of pipeline
stages that can run together as a pipeline of vectorized processing, as
described in Section~\ref{sec:pipelined} of the Appendix;
\texttt{AggregationJobStage}, to perform aggregation on shuffled data that is
generated by a \texttt{PipelineJobStage}; and \texttt{BuildHashTableJobStage}, to build hash
tables using shuffled data or broadcasted data that is generated by a
\texttt{PipelineJobStage}. Then it
dynamically schedules those \texttt{JobStage}s as well as initiates and
monitors execution of the \texttt{JobStage}s on the cluster.
\end{enumerate}

\noindent 
Each worker node in a PC cluster runs two processes: the \emph{worker front-end process} and the \emph{worker backend process}.
Dual processes are used because PC executes potentially unsafe native user code.  By definition, user code is run only in the backend process so that the
worker front-end process is ``crash proof''---if            
user code happens to crash the worker backend process, the worker front-end process will receive a signal and it can re-fork the worker
backend process.

The worker front-end process runs following components:

\begin{enumerate}
\item The \emph{local catalog manager} that requests and 
buffers data served from the master server's catalog manager.  The local catalog manager is also responsible for fetching and dynamically loading
code as needed when a virtual method call is made over a PC \texttt{Object} (see Section \ref{sec:dyn_dis} of the paper).
\item The \emph{local storage server}, that manages a
  shared-memory buffer pool used for buffering and caching datasets.  It also manages
a  user-level file system that is used to persist a portion of one or more datasets stored by PC (the partitioning of datasets to storage servers is managed
by the master server's distributed storage manager).  The storage server also manages temporary data that must be spilled to secondary storage.
The shared memory buffer pool is created via a \texttt{mmap} system call so that
data stored in it can be read by the backend process (forked from
the front-end process) via zero-copy access. Compared with distributed
computation framework such as Spark~\cite{zaharia2010spark} and Flink~\cite{alexandrov2014stratosphere, carbone2015apache}, which rely on
external storage systems such as HDFS~\cite{borthakur2008hdfs} and Alluxio~\cite{li2014tachyon}, a significant
performance advantage of having a buffer pool is that data can be cached in memory
across different applications, rather than having to be load from external
storages (which often requires data movement) each time.
\item And finally, the \emph{message proxy} which is responsible for communicating with the worker backend process, and which acts as a bridge between the
worker backend process and the master server, the local catalog, and the local storage server.

\end{enumerate}

Because worker backend processes are the only processes that are actually allowed to run user code (and because user code is that code that actually processes
user-supplied data) it means that the worker backend processes are
where all computations in PC are actually performed.  One or more computations in PC correspond 
to a set of the 
\texttt{JobStage}s, that are created by the master server's distributed query scheduler, 
for executing the computations and required communications
that link together the various \texttt{JobStage}s to implement distributed
computations such as joins and aggregations. 
For example, considering a query graph that consists of a \texttt{Reader}, a standalone \texttt{SelectionComp} following
the \texttt{Reader}, and a
\texttt{Writer} following the \texttt{SelectionComp}. All of those Computations correspond
to one \texttt{PipelineJobStage}. In the next two subsections, we
describe how PC implements two of its distributed computations:
distributed aggregation, and a distributed hash-partition join.

\subsection{Distributed Aggregation}

\begin{figure}
\centering
\includegraphics[width=0.75\textwidth]{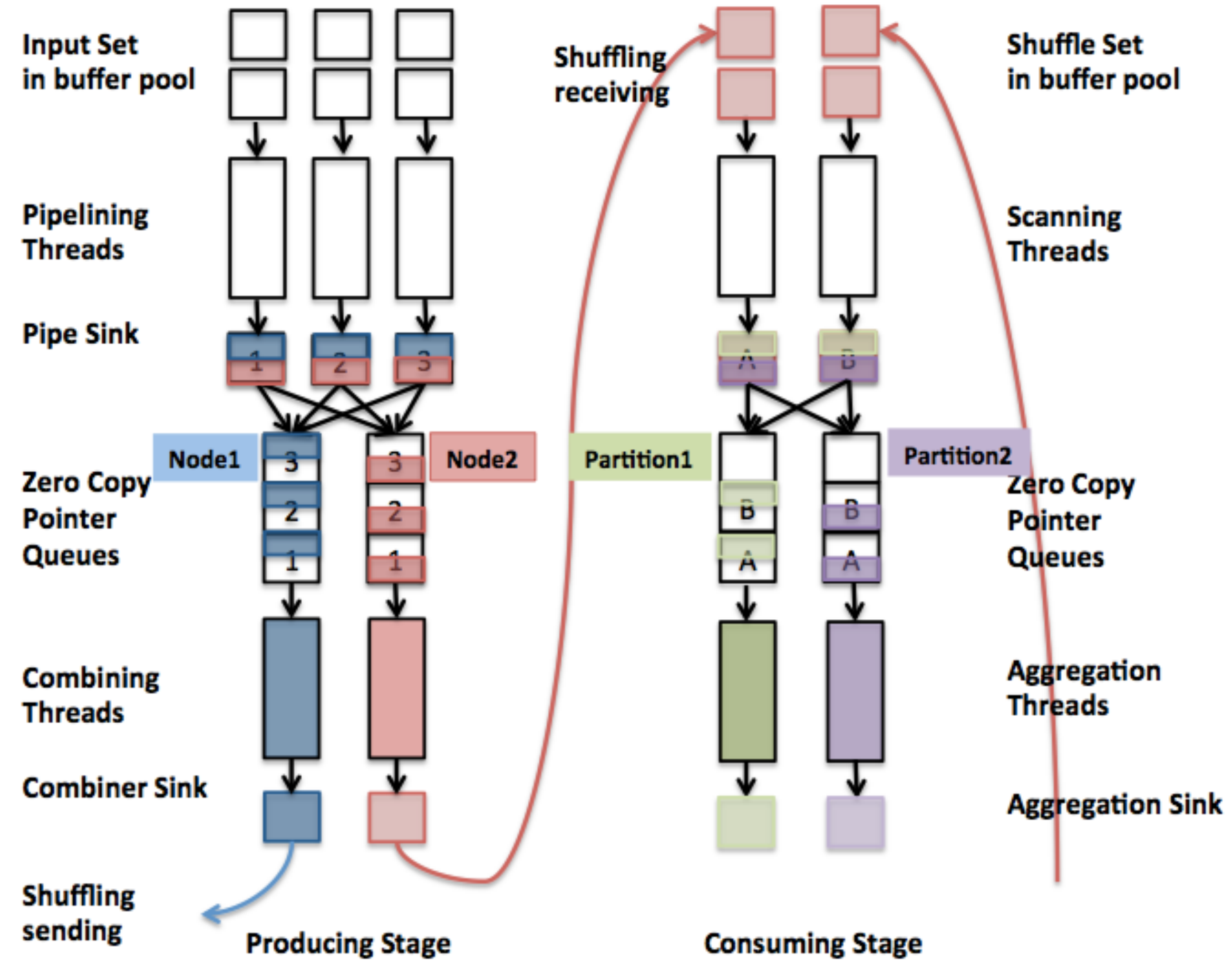}
  \caption{\label{fig:aggregation} Distributed aggregation in PC.}
\end{figure}

The workflow of PC's distributed aggregation implementation is
shown in Figure ~\ref{fig:aggregation}.
The most unique aspect of distributed aggregation in PC is the way in which it is tightly coupled with the PC \texttt{Object} model.  All data structures used
to power distributed aggregation are themselves PC \texttt{Objects},
and hence they all provide efficient data shuffling with zero
serialization and deserialization costs.

Distributed aggregation in PC is broken into two different job stages.
In the first, there is a \texttt{PipelineJob Stage} working as a \emph{producing stage},
where all required pre-aggregation computations are performed, and then the
data are pre-aggregated and stored in a set of hash tables (that is, PC \texttt{Map} objects).  These PC \texttt{Map} objects are shuffled, so that all partial aggregates
associated with the same key are on the same machine.  Then in a
second job stage, which is a \texttt{AggregationJobStage} working as the \emph{consuming stage},
the shuffled \texttt{Map} objects from around the cluster are then aggregated
to produce the final aggregation result.

\vspace{5pt}
{\bf 1. The producing stage.} In this stage, a sequence of TCAP
operations that end in an \texttt{AGGREGATE} operation are used to
create \texttt{PipelineJobStage}, which is 
realized as a list
of pipelined stages, where each stage corresponds to a TCAP operation.
A replica of the list of pipelined stages is
created for each of the $N$ so-called \emph{pipelining threads} dedicated to implementing
the producing stage on each of the backend worker processes.  
The thread assigned to each of those pipelines begins by allocating a PC \texttt{Vector <Handle <Map <Object, Object}\texttt{>}\texttt{>}\texttt{>} on an output page.  This page will serve
as the \emph{pipe sink}.  The pipe sink is
the location where the data produced by the pipeline are stored.  In
this way the PC object model is used to store the results
from the producing stage, so that they can be sent over the network and used at the other side with no serialization/deserialization.

Once the pipe sinks have been allocated,
the system begins assigning input pages to each of the pipelining 
threads.  Each input page was itself produced as the result of a
previous \texttt{JobStage}, or else created by users, both of which needs to be
loaded from the storage.  Hence all pages are themselves organized using the PC object model.  Thus, each input
page will itself contain a \texttt{Vector <Handle <Object}\texttt{>}\texttt{>},
or another standard PC container type.  During the producing stage, vectors of PC \texttt{Handle}s are repeatedly
loaded from each input page, and each is used to create a vector list that is pushed through the list of pipeline stages.

Ultimately, pushing a vector list through the pipeline stages results in an output vector list that contains a vector of keys and a vector of values 
associated with those keys that need to be aggregated.  
For each (key, value)
pair, first the key is hashed to a particular \emph{hash partition}---the hash partition determines which of the \texttt{Map}
objects in the \texttt{Vector <Handle <Map <Object, Object}\texttt{>}\texttt{>}\texttt{>} stored on the output page will receive the pair, and which machine will house the final aggregation
for all instances of the key.  Once the hash partition has been determined,
the pair is added to the associated \texttt{Map} in the pipe sink
(if a particular key is already there, the
existing value is added to the new value, and the result is used to replace the existing value).  

At the same time, 
$K$ different \emph{combining threads} are running in each backend worker process.  
When a page containing a \texttt{Vector <Handle <Map <Object,
  Object}\texttt{>}\texttt{>}\texttt{>} becomes full, the thread running the associated
\texttt{PipelineJobStage} allocates a new pipe sink,
and the filled page is added to the queue of completed pages to be processed.  There is one such queue
associated with each of the $K$ combining threads.
We call each such queue a \emph{zero copy pointer queue} because the entire page of PC \texttt{Object}s is passed from a pipelining thread to each of the
combining threads as a pointer, with no PC \texttt{Object} movement.

Each combining thread is assigned one or more hash partitions.  When a combining thread
is associated with a hash partition, it is responsible for 
aggregating all of the \texttt{Map}s produced by local pipelining threads that were associated with that hash partition.
The result is a 
\texttt{Map} containing data for that one hash partition.  Since each combining
thread can be assigned more than one hash partition that will be sent to the same destination machine,
the combining thread also produces a \texttt{Vector <Handle <Map <Object, Object}\texttt{>}\texttt{>}\texttt{>} that is referred to as a \emph{combining sink}.
Each \texttt{Map} in the \texttt{Vector} is associated with one hash partition.
This object is stored on a thread-local \emph{combiner page} whose size is automatically tuned by the system.

\vspace{5pt}
{\bf 2. The consuming Stage.}
Once a combiner page becomes full, the page is directly sent to the worker node housing the
backend worker processes tasked with processing the set of hash partitions on the page.
This communication is referred to as a \emph{shuffle}.  
Compression is used to reduce the shuffle data size and save on network cost.
When a worker node receives a page during the shuffle, one of a pool of \emph{scanning threads} will look at the 
\texttt{Vector <Handle <Map <Object, Object}\texttt{>}\texttt{>}\texttt{>} object contained in the page, and then add a pointer to the page
to each of the queues associated with the $M$ \emph{aggregation threads}.  These are the threads
that are tasked with final processing of all of the various hash partitions.
Each aggregation thread is responsible for aggregating exactly 
one hash partition, and so $M$ is the total number of hash partitions processed by a given backend worker process.
An aggregation thread has an output page housing its \emph{aggregation sink}.  The aggregation sink is a \texttt{Map <Object, Object>} that stores the portion of the
data that it aggregates.

In PC, all of the ``magic'' parameters such as the number of hash partitions $M$, 
the number of pipelining threads $N$, 
the number of combining threads for each remote node $K$, 
the combiner page size, and the aggregation page size are all automatically 
determined at run-time to maximize memory utilization, network
utilization and CPU utilization. The details are omitted here and will
be discussed in a separate paper.

\subsection{Hash Partition Join}

We now briefly describe how PC implements a distributed equi-join of $n$ different sets using a hash-partition strategy.  
That is, imagine that $t$ = $\langle t_1, t_2, ..., t_n \rangle$ is a
tuple formed by taking one item from each of the $n$ sets.  Let $key(t_i)$ denote the join key of $t_i$.  Then an $n$-way equi-join requires that
in order for $t$ to be in the output set, it must be the case that
$key(t_i) = key(t_j)$ forall $i, j$. 

In PC, such an operation is broken into $2n$ \texttt{JobStage}s. The first $n$
\texttt{JobStage}s are \texttt{PipelineJobStage}s hashing and repartitioning each of the $n$ sets, so that all of the data with the same join key value
must be co-located on the same machine.  The next $n - 1$ \texttt{JobStage}s
are \texttt{BuildHashTableJobStage}s that build hash tables for all of the
entries in $n - 1$ of those sets.  The last \texttt{JobStage} is also a 
\texttt{PipelineJobStage} that scans the last set and
probes the constructed hash tables.  

In more detail, the three types of stages are:

\vspace{5pt}
{\bf 1. The data repartition stages.} This class of \texttt{JobStage}s are
\texttt{PipelineJobStage}s. It is similar to the producing stage in distributed
aggregation, with one key difference.
Rather than using a \texttt{Vector <Handle <Map <Object, Object}\texttt{>}\texttt{>}\texttt{>} to store (key, value) pairs where the value is the result obtained by aggregating a set of
data with the same key, the pipe sink used is instead a \texttt{Vector <Handle <Map <unsigned\_t, Vector <Object}\texttt{>}\texttt{>}\texttt{>}\texttt{>} data structure.  Here, the
\texttt{unsigned\_t} is a hash value produced by a TCAP \texttt{HASH} operation over the input object's join key, and the \texttt{Vector <Object>} is a list of objects
with the same hash value.  
When a new object with the same hash key is found during pipelined processing or during combining, rather than aggregating, the new object is instead inserted
into the inner \texttt{Vector <Object>} data structure that contains a set of objects with the same hash key.
Note that after the data repartition job stages completes, all of the data from all of the sets will have been repartitioned, so that all of the data with the same join
key will be co-located on the same machine.

\vspace{5pt}
{\bf 2. The hash table building stages.} These \texttt{JobStage}s are
\texttt{BuildHashTableJobStage}s, which are similar to the consuming job stage in aggregation, 
except that again, rather than aggregating, the goal is to build up \texttt{Vector}s of objects, stored in various \texttt{Map} data structures (one for each
aggregation thread), where the \texttt{Vector} of objects associated with a particular 
\texttt{unsigned\_t} contains \emph{all} of the objects in a set whose join key hashes to that particular \texttt{unsigned\_t} value.
As a result of this class of \texttt{JobStage}s, the contents of $n - 1$ of the input sets will be stored precisely as required.

\vspace{5pt}
{\bf 3. The hash join stage.} 
This \texttt{JobStage} is also a \texttt{PipelineJobStage}. It runs over the last set (the one that was not 
hashed as part of join stage 2), and output results to the next
\texttt{JobStage} (through shuffling) or an Output sink (e.g. write to storage
as final results). Now imagine that we are processing the last dataset, after shuffling.  At this point,
all of the other sets have been
stored in \texttt{Map <unsigned\_t, Vector <Object}\texttt{>}\texttt{>} objects.  As we process the final set,
the hash value for each object is used to probe the \texttt{Map}
associated with each of the other $n-1$ sets.  If a match is found in each one of those other $n-1$ sets, then
one or more entries in the output vector list are created to store the matches.  The resulting vector sets are then pushed through a pipeline that post-processes
the data in the vector list, likely checking to see if this was an actual match (and not just an accidental hash collision) and perhaps performing the processing
necessary to prepare for the \emph{next} join or aggregation.
Note that if $m_i$ objects from the $i$th input set have the same \texttt{unsigned\_t} hash value, then $\prod_i m_i$ entries in a vector set will be created
as a result of these hash table probes. 

%
%
%

%% file: main.bbl
\begin{thebibliography}{10}

\bibitem{dj4j}
Dl4j.
\newblock \url{https://deeplearning4j.org/}.
\newblock Accessed Sept 27, 2017.

\bibitem{eigen}
Eigen.
\newblock \url{http://eigen.tuxfamily.org/index.php?title=Main_Page}.
\newblock Accessed Oct 22, 2016.

\bibitem{gsl}
Gsl - gnu scientific library.
\newblock \url{https://www.gnu.org/software/gsl/}.
\newblock Accessed Oct 22, 2016.

\bibitem{mahout}
Mahout.
\newblock \url{http://mahout.apache.org}.
\newblock Accessed Oct 22, 2016.

\bibitem{samsara}
Mahout samsara.
\newblock
  \url{https://mahout.apache.org/users/environment/out-of-core-reference.html}.
\newblock Accessed Oct 22, 2016.

\bibitem{tungsten}
Project tungsten: Bringing spark closer to bare metal.
\newblock
  \url{https://databricks.com/blog/2015/04/28/project-tungsten-bringing-spark-closer-to-bare-metal.html}.
\newblock Accessed Nov 1, 2017.

\bibitem{SciDBOS}
Scidb's supported os.
\newblock \url{http://www.paradigm4.com/HTMLmanual/14.8/scidb_ug/pr01s01.html}.
\newblock Accessed Nov 1, 2017.

\bibitem{abadi2009column}
D.~J. Abadi, P.~A. Boncz, and S.~Harizopoulos.
\newblock Column-oriented database systems.
\newblock {\em Proceedings of the VLDB Endowment}, 2(2):1664--1665, 2009.

\bibitem{abadi2016tensorflow}
M.~Abadi, A.~Agarwal, P.~Barham, E.~Brevdo, Z.~Chen, C.~Citro, G.~S. Corrado,
  A.~Davis, J.~Dean, M.~Devin, et~al.
\newblock Tensorflow: Large-scale machine learning on heterogeneous distributed
  systems.
\newblock {\em arXiv preprint arXiv:1603.04467}, 2016.

\bibitem{ahmad2009dbtoaster}
Y.~Ahmad and C.~Koch.
\newblock Dbtoaster: A sql compiler for high-performance delta processing in
  main-memory databases.
\newblock {\em Proceedings of the VLDB Endowment}, 2(2):1566--1569, 2009.

\bibitem{alexandrov2014stratosphere}
A.~Alexandrov et~al.
\newblock The {S}tratosphere platform for big data analytics.
\newblock {\em VLDBJ}, 23(6):939--964, 2014.

\bibitem{alexandrov2015implicit}
A.~Alexandrov, A.~Kunft, A.~Katsifodimos, F.~Sch{\"u}ler, L.~Thamsen, O.~Kao,
  T.~Herb, and V.~Markl.
\newblock Implicit parallelism through deep language embedding.
\newblock In {\em Proceedings of the 2015 ACM SIGMOD International Conference
  on Management of Data}, pages 47--61. ACM, 2015.

\bibitem{armbrust2015spark}
M.~Armbrust, R.~S. Xin, C.~Lian, Y.~Huai, D.~Liu, J.~K. Bradley, X.~Meng,
  T.~Kaftan, M.~J. Franklin, A.~Ghodsi, et~al.
\newblock Spark sql: Relational data processing in spark.
\newblock In {\em Proceedings of the 2015 ACM SIGMOD International Conference
  on Management of Data}, pages 1383--1394. ACM, 2015.

\bibitem{bacon2017spanner}
D.~F. Bacon, N.~Bales, N.~Bruno, B.~F. Cooper, A.~Dickinson, A.~Fikes,
  C.~Fraser, A.~Gubarev, M.~Joshi, E.~Kogan, et~al.
\newblock Spanner: Becoming a sql system.
\newblock In {\em Proceedings of the 2017 ACM International Conference on
  Management of Data}, pages 331--343. ACM, 2017.

\bibitem{barendregt1984lambda}
H.~P. Barendregt et~al.
\newblock {\em The lambda calculus}, volume~2.
\newblock North-Holland Amsterdam, 1984.

\bibitem{bittorf2015impala}
M.~Bittorf, T.~Bobrovytsky, C.~C. A. C.~J. Erickson, M.~G.~D. Hecht, M.~J. I.
  J.~L. Kuff, D.~K.~A. Leblang, N.~L. I. P.~H. Robinson, D.~R.~S. Rus, J.~R. D.
  T.~S. Wanderman, and M.~M. Yoder.
\newblock Impala: A modern, open-source sql engine for hadoop.
\newblock In {\em Proceedings of the 7th Biennial Conference on Innovative Data
  Systems Research}, 2015.

\bibitem{blackburn2006dacapo}
S.~M. Blackburn, R.~Garner, C.~Hoffmann, A.~M. Khang, K.~S. McKinley,
  R.~Bentzur, A.~Diwan, D.~Feinberg, D.~Frampton, S.~Z. Guyer, et~al.
\newblock The dacapo benchmarks: Java benchmarking development and analysis.
\newblock In {\em ACM Sigplan Notices}, volume~41, pages 169--190. ACM, 2006.

\bibitem{blei2003latent}
D.~M. Blei, A.~Y. Ng, and M.~I. Jordan.
\newblock Latent dirichlet allocation.
\newblock {\em Journal of machine Learning research}, 3(Jan):993--1022, 2003.

\bibitem{blumofe1996cilk}
R.~D. Blumofe, C.~F. Joerg, B.~C. Kuszmaul, C.~E. Leiserson, K.~H. Randall, and
  Y.~Zhou.
\newblock Cilk: An efficient multithreaded runtime system.
\newblock {\em Journal of parallel and distributed computing}, 37(1):55--69,
  1996.

\bibitem{boehm2016systemml}
M.~Boehm, M.~W. Dusenberry, D.~Eriksson, A.~V. Evfimievski, F.~M. Manshadi,
  N.~Pansare, B.~Reinwald, F.~R. Reiss, P.~Sen, A.~C. Surve, et~al.
\newblock Systemml: Declarative machine learning on spark.
\newblock {\em Proceedings of the VLDB Endowment}, 9(13):1425--1436, 2016.

\bibitem{boehm2014hybrid}
M.~Boehm, S.~Tatikonda, B.~Reinwald, P.~Sen, Y.~Tian, D.~R. Burdick, and
  S.~Vaithyanathan.
\newblock Hybrid parallelization strategies for large-scale machine learning in
  systemml.
\newblock {\em Proceedings of the VLDB Endowment}, 7(7):553--564, 2014.

\bibitem{boncz2005monetdb}
P.~A. Boncz, M.~Zukowski, and N.~Nes.
\newblock Monetdb/x100: Hyper-pipelining query execution.
\newblock In {\em Cidr}, volume~5, pages 225--237, 2005.

\bibitem{borthakur2008hdfs}
D.~Borthakur.
\newblock Hdfs architecture guide.
\newblock {\em HADOOP APACHE PROJECT http://hadoop. apache.
  org/common/docs/current/hdfs design. pdf}, 2008.

\bibitem{bress2017generating}
S.~Bre{\ss}, B.~K{\"o}cher, H.~Funke, T.~Rabl, and V.~Markl.
\newblock Generating custom code for efficient query execution on heterogeneous
  processors.
\newblock {\em arXiv preprint arXiv:1709.00700}, 2017.

\bibitem{brown2010overview}
P.~G. Brown.
\newblock Overview of scidb: large scale array storage, processing and
  analysis.
\newblock In {\em Proceedings of the 2010 ACM SIGMOD International Conference
  on Management of data}, pages 963--968. ACM, 2010.

\bibitem{jermaineExperimental}
Z.~Cai, Z.~J. Gao, S.~Luo, L.~L. Perez, Z.~Vagena, and C.~Jermaine.
\newblock A comparison of platforms for implementing and running very large
  scale machine learning algorithms.
\newblock In {\em Proceedings of the 2014 ACM SIGMOD international conference
  on Management of data}, pages 1371--1382. ACM, 2014.

\bibitem{carbone2015apache}
P.~Carbone, A.~Katsifodimos, S.~Ewen, V.~Markl, S.~Haridi, and K.~Tzoumas.
\newblock Apache flink: Stream and batch processing in a single engine.
\newblock {\em Bulletin of the IEEE Computer Society Technical Committee on
  Data Engineering}, 36(4), 2015.

\bibitem{chaudhuri1998overview}
S.~Chaudhuri.
\newblock An overview of query optimization in relational systems.
\newblock In {\em Proceedings of the seventeenth ACM SIGACT-SIGMOD-SIGART
  symposium on Principles of database systems}, pages 34--43. ACM, 1998.

\bibitem{chen1993hilog}
W.~Chen, M.~Kifer, and D.~S. Warren.
\newblock Hilog: A foundation for higher-order logic programming.
\newblock {\em The Journal of Logic Programming}, 15(3):187--230, 1993.

\bibitem{codd1971data}
E.~F. Codd.
\newblock A data base sublanguage founded on the relational calculus.
\newblock In {\em Proceedings of the 1971 ACM SIGFIDET (now SIGMOD) Workshop on
  Data Description, Access and Control}, pages 35--68. ACM, 1971.

\bibitem{council2008tpc}
T.~P.~P. Council.
\newblock Tpc-h benchmark specification.
\newblock {\em Published at http://www. tcp. org/hspec. html}, 21:592--603,
  2008.

\bibitem{crotty2015tupleware}
A.~Crotty, A.~Galakatos, K.~Dursun, T.~Kraska, U.~Cetintemel, and S.~B. Zdonik.
\newblock Tupleware:" big" data, big analytics, small clusters.
\newblock In {\em CIDR}, 2015.

\bibitem{dagum1998openmp}
L.~Dagum and R.~Menon.
\newblock Openmp: an industry standard api for shared-memory programming.
\newblock {\em IEEE computational science and engineering}, 5(1):46--55, 1998.

\bibitem{dave2016graphframes}
A.~Dave, A.~Jindal, L.~E. Li, R.~Xin, J.~Gonzalez, and M.~Zaharia.
\newblock Graphframes: an integrated api for mixing graph and relational
  queries.
\newblock In {\em Proceedings of the Fourth International Workshop on Graph
  Data Management Experiences and Systems}, page~2. ACM, 2016.

\bibitem{dean2008mapreduce}
J.~Dean and S.~Ghemawat.
\newblock Mapreduce: Simplified data processing on large clusters.
\newblock {\em Communications of the ACM}, 51(1):107--113, 2008.

\bibitem{ghoting2011systemml}
A.~Ghoting, R.~Krishnamurthy, E.~Pednault, B.~Reinwald, V.~Sindhwani,
  S.~Tatikonda, Y.~Tian, and S.~Vaithyanathan.
\newblock System{ML}: Declarative machine learning on mapreduce.
\newblock In {\em ICDE}, pages 231--242, 2011.

\bibitem{gonzalez2014graphx}
J.~E. Gonzalez, R.~S. Xin, A.~Dave, D.~Crankshaw, M.~J. Franklin, and
  I.~Stoica.
\newblock Graphx: Graph processing in a distributed dataflow framework.
\newblock In {\em OSDI}, volume~14, pages 599--613, 2014.

\bibitem{graefe1990encapsulation}
G.~Graefe.
\newblock {\em Encapsulation of parallelism in the Volcano query processing
  system}, volume~19.
\newblock ACM, 1990.

\bibitem{graefe1995cascades}
G.~Graefe.
\newblock The cascades framework for query optimization.
\newblock {\em IEEE Data Eng. Bull.}, 18(3):19--29, 1995.

\bibitem{gropp1996high}
W.~Gropp, E.~Lusk, N.~Doss, and A.~Skjellum.
\newblock A high-performance, portable implementation of the mpi message
  passing interface standard.
\newblock {\em Parallel computing}, 22(6):789--828, 1996.

\bibitem{grossman2002region}
D.~Grossman, G.~Morrisett, T.~Jim, M.~Hicks, Y.~Wang, and J.~Cheney.
\newblock Region-based memory management in cyclone.
\newblock {\em ACM Sigplan Notices}, 37(5):282--293, 2002.

\bibitem{idreos2012monetdb}
S.~Idreos, F.~Groffen, N.~Nes, S.~Manegold, S.~Mullender, M.~Kersten, et~al.
\newblock Monetdb: Two decades of research in column-oriented database
  architectures.
\newblock {\em A Quarterly Bulletin of the IEEE Computer Society Technical
  Committee on Database Engineering}, 35(1):40--45, 2012.

\bibitem{jarke1984query}
M.~Jarke and J.~Koch.
\newblock Query optimization in database systems.
\newblock {\em ACM Computing surveys (CsUR)}, 16(2):111--152, 1984.

\bibitem{josuttis2012c++}
N.~M. Josuttis.
\newblock {\em The C++ standard library: a tutorial and reference}.
\newblock Addison-Wesley, 2012.

\bibitem{kale1993charm++}
L.~V. Kale and S.~Krishnan.
\newblock Charm++: a portable concurrent object oriented system based on c++.
\newblock In {\em ACM Sigplan Notices}, volume~28, pages 91--108. ACM, 1993.

\bibitem{klonatos2014building}
Y.~Klonatos, C.~Koch, T.~Rompf, and H.~Chafi.
\newblock Building efficient query engines in a high-level language.
\newblock {\em Proceedings of the VLDB Endowment}, 7(10):853--864, 2014.

\bibitem{lattner2004llvm}
C.~Lattner and V.~Adve.
\newblock Llvm: A compilation framework for lifelong program analysis \&
  transformation.
\newblock In {\em Proceedings of the international symposium on Code generation
  and optimization: feedback-directed and runtime optimization}, page~75. IEEE
  Computer Society, 2004.

\bibitem{lattner2002llvm}
C.~A. Lattner.
\newblock {\em LLVM: An infrastructure for multi-stage optimization}.
\newblock PhD thesis, University of Illinois at Urbana-Champaign, 2002.

\bibitem{lebanon2006metric}
G.~Lebanon.
\newblock Metric learning for text documents.
\newblock {\em IEEE Transactions on Pattern Analysis and Machine Intelligence},
  28(4):497--508, 2006.

\bibitem{li2014tachyon}
H.~Li and et~al.
\newblock Tachyon: Reliable, memory speed storage for cluster computing
  frameworks.
\newblock In {\em SOCC}, pages 1--15, 2014.

\bibitem{lu2016lifetime}
L.~Lu and et~al.
\newblock Lifetime-based memory management for distributed data processing
  systems.
\newblock {\em VLDB}, 9(12):936--947, 2016.

\bibitem{meng2016mllib}
X.~Meng, J.~Bradley, B.~Yavuz, E.~Sparks, S.~Venkataraman, D.~Liu, J.~Freeman,
  D.~Tsai, M.~Amde, S.~Owen, et~al.
\newblock Mllib: Machine learning in apache spark.
\newblock {\em The Journal of Machine Learning Research}, 17(1):1235--1241,
  2016.

\bibitem{miller1991logic}
D.~Miller.
\newblock A logic programming language with lambda-abstraction, function
  variables, and simple unification.
\newblock {\em Journal of logic and computation}, 1(4):497--536, 1991.

\bibitem{moggi1989computational}
E.~Moggi.
\newblock Computational lambda-calculus and monads.
\newblock In {\em Logic in Computer Science, 1989. LICS'89, Proceedings.,
  Fourth Annual Symposium on}, pages 14--23. IEEE, 1989.

\bibitem{nagel2014code}
F.~Nagel, G.~Bierman, and S.~D. Viglas.
\newblock Code generation for efficient query processing in managed runtimes.
\newblock {\em Proceedings of the VLDB Endowment}, 7(12):1095--1106, 2014.

\bibitem{neumann2011efficiently}
T.~Neumann.
\newblock Efficiently compiling efficient query plans for modern hardware.
\newblock {\em Proceedings of the VLDB Endowment}, 4(9):539--550, 2011.

\bibitem{newburn2011intel}
C.~J. Newburn, B.~So, Z.~Liu, M.~McCool, A.~Ghuloum, S.~D. Toit, Z.~G. Wang,
  Z.~H. Du, Y.~Chen, G.~Wu, et~al.
\newblock Intel's array building blocks: A retargetable, dynamic compiler and
  embedded language.
\newblock In {\em Proceedings of the 9th Annual IEEE/ACM International
  Symposium on Code Generation and Optimization}, pages 224--235. IEEE Computer
  Society, 2011.

\bibitem{ousterhout2015making}
K.~Ousterhout, R.~Rasti, S.~Ratnasamy, S.~Shenker, B.-G. Chun, and V.~ICSI.
\newblock Making sense of performance in data analytics frameworks.
\newblock In {\em NSDI}, volume~15, pages 293--307, 2015.

\bibitem{palkar2017weld}
S.~Palkar, J.~J. Thomas, A.~Shanbhag, D.~Narayanan, H.~Pirk, M.~Schwarzkopf,
  S.~Amarasinghe, M.~Zaharia, and S.~InfoLab.
\newblock Weld: A common runtime for high performance data analytics.
\newblock In {\em Conference on Innovative Data Systems Research (CIDR)}, 2017.

\bibitem{pirk2016voodoo}
H.~Pirk, O.~Moll, M.~Zaharia, and S.~Madden.
\newblock Voodoo-a vector algebra for portable database performance on modern
  hardware.
\newblock {\em Proceedings of the VLDB Endowment}, 9(14):1707--1718, 2016.

\bibitem{reinders2007intel}
J.~Reinders.
\newblock {\em Intel threading building blocks: outfitting C++ for multi-core
  processor parallelism}.
\newblock " O'Reilly Media, Inc.", 2007.

\bibitem{shi2015clash}
J.~Shi, Y.~Qiu, U.~F. Minhas, L.~Jiao, C.~Wang, B.~Reinwald, and F.~{\"O}zcan.
\newblock Clash of the titans: Mapreduce vs. spark for large scale data
  analytics.
\newblock {\em Proceedings of the VLDB Endowment}, 8(13):2110--2121, 2015.

\bibitem{SikdarSoCC2017}
S.~Sikdar, K.~Teymourian, and C.~Jermaine.
\newblock An experimental comparison of complex object implementations for big
  data systems.
\newblock In {\em Proceedings of the 2017 Symposium on Cloud Computing}, SoCC
  '17, pages 432--444. ACM, 2017.

\bibitem{stonebraker2005c}
M.~Stonebraker, D.~J. Abadi, A.~Batkin, X.~Chen, M.~Cherniack, M.~Ferreira,
  E.~Lau, A.~Lin, S.~Madden, E.~O'Neil, et~al.
\newblock C-store: a column-oriented dbms.
\newblock In {\em Proceedings of the 31st international conference on Very
  large data bases}, pages 553--564. VLDB Endowment, 2005.

\bibitem{stonebraker2011architecture}
M.~Stonebraker, P.~Brown, A.~Poliakov, and S.~Raman.
\newblock The architecture of scidb.
\newblock In {\em Scientific and Statistical Database Management}, pages 1--16.
  Springer, 2011.

\bibitem{stonebraker1990third}
M.~Stonebraker, L.~A. Rowe, B.~G. Lindsay, J.~Gray, M.~J. Carey, M.~L. Brodie,
  P.~A. Bernstein, and D.~Beech.
\newblock Third-generation database system manifesto.
\newblock {\em ACM sIGMOD record}, 19(3):31--44, 1990.

\bibitem{tian2012scalable}
Y.~Tian, S.~Tatikonda, and B.~Reinwald.
\newblock Scalable and numerically stable descriptive statistics in systemml.
\newblock In {\em Data Engineering (ICDE), 2012 IEEE 28th International
  Conference on}, pages 1351--1359. IEEE, 2012.

\bibitem{tofte1997region}
M.~Tofte and J.-P. Talpin.
\newblock Region-based memory management.
\newblock {\em Information and computation}, 132(2):109--176, 1997.

\bibitem{yu2008dryadlinq}
Y.~Yu et~al.
\newblock Dryadlinq: A system for general-purpose distributed data-parallel
  computing using a high-level language.
\newblock In {\em OSDI}, volume~8, pages 1--14, 2008.

\bibitem{zaharia2010spark}
M.~Zaharia, M.~Chowdhury, M.~J. Franklin, S.~Shenker, and I.~Stoica.
\newblock Spark: cluster computing with working sets.
\newblock In {\em USENIX HotCloud}, pages 1--10, 2010.

\bibitem{zaharia2012resilient}
M.~Zaharia and et~al.
\newblock Resilient distributed datasets: A fault-tolerant abstraction for
  in-memory cluster computing.
\newblock In {\em NSDI}, pages 2--15. USENIX, 2012.

\bibitem{zukowski2005monetdb}
M.~Zukowski, P.~A. Boncz, N.~Nes, and S.~H{\'e}man.
\newblock Monetdb/x100-a dbms in the cpu cache.
\newblock {\em IEEE Data Eng. Bull.}, 28(2):17--22, 2005.

\bibitem{zukowski2012vectorwise}
M.~Zukowski, M.~van~de Wiel, and P.~Boncz.
\newblock Vectorwise: A vectorized analytical dbms.
\newblock In {\em Data Engineering (ICDE), 2012 IEEE 28th International
  Conference on}, pages 1349--1350. IEEE, 2012.

\end{thebibliography}
